\documentclass[twocolumn,aps,superscriptaddress,showpacs,nofootinbib,floatfix]{revtex4}
\usepackage{listings}
\usepackage[utf8]{inputenc}
\usepackage{hyperref}
\usepackage{amsmath}
\usepackage{amssymb}
\usepackage{amsfonts}
\usepackage{tabularx}
\usepackage{booktabs}
\usepackage{graphicx}
\usepackage{color}
\usepackage{multirow}
\usepackage[inline]{enumitem}
\usepackage[T2A,T1]{fontenc}

\graphicspath{{fig/}}
\definecolor{theblue}{RGB}{0,50,230}

\usepackage{hyperref}
\hypersetup{
  colorlinks=true,
  linkcolor=theblue,
  citecolor=theblue,
  urlcolor=theblue
}

\newcommand {\avg}[1]{\ensuremath{\langle\kern-1.0pt\langle#1\rangle\kern-1.0pt\rangle}}

\newlength\cmsFigWidth

\usepackage[normalem]{ulem}  % \sout{old text} for strikeout
\usepackage{soul, xcolor}

\renewcommand\sout{\bgroup \color{red} \ULdepth=-.5ex \ULset}
\begin{document}

\title{ Beam Energy dependence of Light Nuclei Production   in  Au+Au Collisions}

\author{Wenbin~Zhao}
\affiliation{Department of Physics and State Key Laboratory of Nuclear Physics and Technology, Peking University, Beijing 100871, China}
\affiliation{Collaborative Innovation Center of Quantum Matter, Beijing 100871, China}
\affiliation{Center for High Energy Physics, Peking University, Beijing 100871, China}
\affiliation{Institute of Particle Physics and Key Laboratory of Quark and Lepton Physics (MOE), Central China Normal University, Wuhan, Hubei 430079, China}
\author{Chun~Shen}
\affiliation{Department of Physics and Astronomy, Wayne State University, Detroit, MI 48201, USA}
\affiliation{RIKEN BNL Research Center, Brookhaven National Laboratory, Upton, NY 11973, USA}

\author{Che~Ming~Ko}
\affiliation{Department of Physics and astronomy, Cyclotron Institute, a Texas A\&M University, College Station, TX 77843, USA}

\author{Quansheng~Liu}
\affiliation{Department of Physics and State Key Laboratory of Nuclear Physics and Technology, Peking University, Beijing 100871, China}
\affiliation{Collaborative Innovation Center of Quantum Matter, Beijing 100871,
China}

\author{Huichao~Song}
\affiliation{Department of Physics and State Key Laboratory of Nuclear Physics and Technology, Peking University, Beijing 100871, China}
\affiliation{Collaborative Innovation Center of Quantum Matter, Beijing 100871, China}
\affiliation{Center for High Energy Physics, Peking University, Beijing 100871, China}

\date{\today}

\begin{abstract}
We study in the coalescence model the collision energy dependence of (anti-)deuteron and (anti-)triton production in the most central Au+Au collisions at $\sqrt{s_{NN}}$ = 7.7, 11.5, 19.6, 27, 39, 62.4 and 200 GeV. The needed phase-space distribution of nucleons at the kinetic freeze-out  is generated from a new 3D hybrid dynamical model (\texttt{iEBE-MUSIC}) by using a smooth  crossover  equation of state (EoS) without a QCD critical point. Our model calculations predict that the coalescence parameters of (anti-)deuteron ($B_2(d)$ and $B_2(\bar{d})$) decrease monotonically as the   collision energy  increases, and the light nuclei yield ratio $N_t  N_p/N_d^2$ remains  approximately  a constant with respect to the  collision energy.   These calculated  observables fail to reproduce the non-monotonic behavior of the corresponding data from the STAR  Collaboration.    Without including  any effects of the critical point in our model, our results serve as  the  baseline predictions for the yields  of light nuclei  in the  search for  the possible QCD critical points from the experimental beam energy scan of heavy ion collisions.
\end{abstract}

\maketitle %Stephanov:2007fk,Asakawa:2015ybt,Luo:2017faz
\section{Introduction}

One of the  primary  goals of  the experiments  at the Relativistic Heavy Ion Collider (RHIC) is to explore and map out the phase structure of  QCD~\cite{Stephanov:2004wx,Stephanov:2007fk,Aoki:2006we,Aggarwal:2010cw,
Fukushima:2010bq,Akiba:2015jwa,Heinz:2015tua,Asakawa:2015ybt,Luo:2017faz,Bzdak:2019pkr}. In particular, the search  for the conjectured critical point in the QCD phase diagram has attracted  much  interest  in the past ten years
~\cite{Asakawa:2015ybt,Luo:2017faz,Bzdak:2019pkr,Stephanov:2008qz,Stephanov:2011pb,Athanasiou:2010kw,Asakawa:2009aj,
Nahrgang:2011mg,Ling:2015yau,Jiang:2015hri,Jiang:2017mji,Mukherjee:2015swa,Mukherjee:2016kyu,
Stephanov:2017ghc,Sakaida:2017rtj,Borsanyi:2018grb,Akamatsu:2018vjr,Nahrgang:2018afz,Odyniec:2019kfh}.   Experiments at  the RHIC Beam Energy Scan (BES) program have already  found some intriguing    results  that might be related to the critical phenomenon in  QCD matter. For example, the cumulant ratio $k\sigma^2$ of  the katosis $\kappa$ and variance $\sigma^2$ of the  (net) proton  multiplicity distribution  obviously deviates from the Poisson  distribution expected from statistical fluctuations  and shows a non-monotonic behavior at lower collision energies~\cite{Luo:2015ewa}. Also, the  Gaussian emission source radii difference ($R^2_{out}-R^2_{side}$) extracted from two-pion interferometry measurements is  found  to have a non-monotonic dependence on the collision energy with a maximum value at around $\sqrt{s_\mathrm{NN}}=$ 20-40 GeV~\cite{Aamodt:2011mr,Adamczyk:2014mxp,Lacey:2014wqa}. Furthermore, the measured yield ratio $N_t  N_p/N_d^2$ of proton, deuteron and triton in central Au+Au collisions  clearly shows a  non-monotonic  behavior in its   collision energy  dependence  with a peak around $\sqrt{s_\mathrm{NN}}=$ 20 GeV~\cite{Zhang:2019wun}.

Besides studying the signatures of  critical fluctuations  in heavy ion collisions, it is also  important and necessary to systematically investigate and  understand  the noncritical and/or thermal fluctuations  that are present in these collisions as they  provide the background against which the signals can be identified   and used  to locate the position of  the possible  critical point  in the QCD phase diagram~\cite{Asakawa:2015ybt,Luo:2017faz,Karsch:2010ck,Borsanyi:2013hza,Luo:2014tga,
Netrakanti:2014mta,Li:2017via,He:2016uei,Xu:2016qjd,Liu:2019nii}.
However,  because of the many complicated processes involved in  realistic heavy-ion collisions,  it is difficult to obtain  clean baseline contributions to   observables in these collisions. For example,  the net-proton multiplicity distribution, which has been suggested as a sensitive signal for  the QCD critical point~\cite{Stephanov:2008qz,Stephanov:2011pb,Athanasiou:2010kw}, is strongly influenced by  both volume fluctuations and charge conservations, which  result in deviations  from the  Skellam  distribution~\cite{Borsanyi:2013hza,Luo:2014tga,Netrakanti:2014mta,Li:2017via}.  To impose strict charge conservations in the hybrid model simulations for QGP and hadronic evolution turns out to be difficult because the local correlation length between  a charged particle  pair   is finite and is sensitive to the   expansion of the  produced  fireball~\cite{Pratt:2018ebf, Oliinychenko:2019zfk}. It is thus highly nontrivial to include all of the important effects originated from non-critical fluctuations in a single model and calculate their contributions to  the higher-order cumulants and the  cumulant ratio of net-proton multiplicity distribution.
%~\cite{Stephanov:2008qz,Athanasiou:2010kw}.

Recently, the STAR Collaboration has collected  a  wealth  of  data on  light nuclei, such as (anti-)deuteron ($\bar{d}$, $d$), (anti-)triton ($\bar{t}$, $t$) and (anti-)helium-3 ($^3\bar{H}e$, $^3He$), and has also  analyzed the energy dependence of  their  yields and  yield ratios in heavy ion collisions  at RHIC BES energies~\cite{Adamczyk:2016gfs,Chen:2018tnh,Adam:2019wnb,Zhang:2019wun}. The  observed coalescence parameters of (anti-)deuteron ($B_2(d)$ and $B_2(\bar{d})$) and the yield ratio of light nuclei, $N_t N_p/N_d^2$,  both  show  a  clear non-monotonic energy dependence with a dip and a peak around $\sqrt{s_\mathrm{NN}}=$20 GeV in central Au+Au collisions, respectively~\cite{Adam:2019wnb,Zhang:2019wun}, implying a  dramatic change of the speed of sound and a  large relative density fluctuations of nucleons   associated with the QCD critical point~\cite{Sun:2017xrx,Sun:2018jhg,Yu:2018kvh,Liu:2019nii,Sun:2020uoj,Deng:2020zxo}. For a better understanding of these observables and evaluate their relations to critical behaviors, it is  necessary and timely to  carry out  baseline calculations without including any effects  from  critical fluctuations.

In this paper, we study the  collision  energy  dependence of  light nuclei production at the RHIC BES  energies  based on  the  nucleon coalescence model using the nucleon phase-space distributions  that do not contain any critical fluctuation effects. More specifically,  nucleons are first  thermally produced  and evolved to  the kinetic freeze-out of an  expanding fireball  described by the integrated hybrid approach \texttt{iEBE-MUSIC} with dynamical initial  conditions that have  been specifically  developed for heavy ion collisions at the RHIC BES program.~\cite{Shen:2017bsr, Shen:2017fnn, 3DGlauber, Denicol:2018wdp, Shen:2018pty}.  With the obtained  phase-space distributions of  protons and neutrons, we then implement the nucleon  coalescence model to calculate the yields of light nuclei~\cite{Mattiello:1995xg,Mattiello:1996gq,Chen:2003qj,Chen:2003ava}. Compared to  previous  studies  based on the thermal model or a transport model without the partonic phase~\cite{Sun:2017xrx,Sun:2018jhg,Yu:2018kvh,Liu:2019nii,Sun:2020uoj},  our present
hybrid model provides a more realistic calculation for light nuclei production  without the effect of the QCD critical point, which  can thus  serve  as more reliable baseline results  for the related measurements  in the experiments carried out  in the RHIC BES program to search for the QCD critical point.

This paper is organized as the following: Section~\ref{sec:model} briefly introduces the nucleon coalescence model and the \texttt{iEBE-MUSIC} hybrid model.   Section~\ref{sec:results} presents and discusses results on the collision energy dependence of the spectra and  yield $dN/dy$ of various hadrons and light nuclei, the coalescence parameters of (anti-)deuterons and (anti-)tritons, and the particle yield ratios in the most central Au+Au collisions at RHIC BES energies.  Section~\ref{sec:summary} concludes the paper.

\section{The theoretical framework}\label{sec:model}

\subsection{The coalescence model for light nuclei production}\label{coalescence}

In the coalescence model~\cite{Mattiello:1995xg,Mattiello:1996gq,Chen:2003qj,Chen:2003ava}, light nuclei are produced  by combining  nucleons at their kinetic freeze-out with probabilities calculated in  the sudden approximation. The production probability for  a (anti-)nucleus of atomic number $A$ consisting of $Z$ (anti-)protons and $N$ (anti-)neuterons ($A=Z+N$)  is given by the overlap of the Wigner function  $f_A$ of the nucleus with the phase-space distributions  $f_{p/\bar{p}}({\bf x}_i, {\bf p}_i, t)$ of (anti-)proton and   $f_{n/\bar{n}}({\bf x}_j, {\bf p}_j, t)$ of (anti-)neutrons~\cite{Chen:2003qj,Chen:2003ava}:
\begin{eqnarray}
\label{coal}
&&\frac{dN_A}{d^3 {\mathbf P}_A}=\frac{g_A}{Z!  N!}\int  \prod_{i=1}^Zp_i^\mu d^3\sigma_{i\mu}\frac{d^3{\bf p}_i}{E_i}f_{p/\bar{p}}({\bf x}_i, {\bf p}_i,t_i)\nonumber\\
&&\times\int  \prod_{j=1}^Np_j^\mu d^3\sigma_{j\mu}\frac{d^3{\bf p}_j}{E_j}f_{n/\bar{n}}({\bf x}_j, {\bf p}_j,t_j)\nonumber\\
&&\times f_A({\bf x}_1^\prime, ... ,{\bf x}_Z^\prime,{\bf x}_1^\prime, ... ,{\bf x}_N^\prime; {\bf p}_1^\prime, ... ,{\bf p}_Z^\prime,{\bf p}_1^\prime, ... ,{\bf p}_N^\prime;t^\prime)\nonumber\\
&&\times \delta^{(3)}\left({\bf P}_A-\sum_{i=1}^Z{\bf p}_i-\sum_{j=1}^N{\bf p}_j\right),
\end{eqnarray}
where $g_A=(2J_A+1)/[\Pi_{i=1}^{A}(2J_i+1)]$ is the statistical factor for $A$ nucleons of spins $J_i$ to form a nucleus of angular momentum $J_A$. The coordinate and momentum of the $i$-th nucleon in the fireball frame are denoted by ${\bf x}_i$ and ${\bf p}_i$, respectively. Its coordinate ${\bf x}_i^\prime$ and momentum ${\bf p}_i^\prime$ in the Wigner function of the produced nucleus are obtained by Lorentz transforming the coordinate ${\bf x}_i$ and momentum ${\bf p}_i$ to the rest frame of the nucleus.

In this paper, we focus on investigating the collision energy dependence of the production  of   (anti-)deuterons and  (anti-)tritons in the RHIC BES program. Following Ref.~\cite{Chen:2003qj},  the Wigner functions of  (anti-)deuterons and  (anti-)tritons are taken  to have  the forms~\cite{Song:2012cd}
\begin{eqnarray}
f_2(\boldsymbol\rho,{\bf p}_\rho)=8\exp\left[-\frac{\boldsymbol\rho^2}{\sigma_\rho^2}-{\bf p}_\rho^2\sigma_\rho^2\right],
\label{two}
\end{eqnarray}
and
\begin{eqnarray}
&&f_3(\boldsymbol\rho,\boldsymbol\lambda,{\bf p}_\rho,{\bf p}_\lambda)\nonumber\\
&&=8^2\exp\left[-\frac{\boldsymbol\rho^2}{\sigma_\rho^2}-\frac{\boldsymbol\lambda^2}{\sigma_\lambda^2}-{\bf p}_\rho^2\sigma_\rho^2-{\bf p}_\lambda^2\sigma_\lambda^2\right],
\label{three}
\end{eqnarray}
respectively.
Here the relative coordinates $\boldsymbol\rho$ and  $\boldsymbol\lambda$, and the  relative  momenta  ${\bf p}_\rho$and ${\bf p}_\lambda$ are defined as:
\begin{eqnarray}\label{rel}
&&\boldsymbol\rho=\frac{1}{\sqrt{2}}({\bf x}_1^\prime-{\bf x}_2^\prime),\quad{\bf p}_\rho=\sqrt{2}~\frac{m_2{\bf p}_1^\prime-m_1{\bf p}_2^\prime}{m_1+m_2},\nonumber\\
&&{\boldsymbol\lambda}=\sqrt{\frac{2}{3}}\left(\frac{m_1{\bf x}_1^\prime+m_2{\bf x}_2^\prime}{m_1+m_2}-{\bf x}_3^\prime\right),\nonumber\\
&&{\bf p}_\lambda=\sqrt{\frac{3}{2}}~\frac{m_3({\bf p}_1^\prime+{\bf p}_2^\prime)-(m_1+m_2){\bf p}_3^\prime}{m_1+m_2+m_3},
\end{eqnarray}
with $m_i$, ${\bf x}_i^\prime$ and ${\bf p}_i^\prime$ being the mass, coordinate and momentum of nucleon $i$, respectively. The width parameter $\sigma_\rho$ in Eq.~(\ref{two}) is related to the root-mean-square charge radius of the nucleus of two constituent  nucleons via~\cite{Song:2012cd}
\begin{eqnarray}\label{r2}
\langle r_2^2 \rangle=\frac{3}{2}\frac{|Q_1m_1^2+Q_2m_2^2|}{(m_1+m_2)^2} \sigma_\rho^2=\frac{3}{4}\frac{ |Q_1 m_1^2+ Q_2 m_2^2 | }{\omega  m_1m_2(m_1+m_2)} \end{eqnarray}
 with $Q_1$ and $Q_2$ being the charges of the two nucleons,  which provides the relation $\sigma_\rho=1/\sqrt{\mu_1\omega}$ in terms of the oscillator frequency $\omega$ in the harmonic wave function and the reduced mass $\mu_1=2(1/m_1+1/m_2)^{-1}$.
The width parameter $\sigma_\lambda$ in Eq.~(\ref{three}) is related to the oscillator frequency by $\sigma_\lambda=1/\sqrt{\mu_2 \omega}$  with $\mu_2=(3/2)[1/(m_1+m_2)+1/m_3]^{-1}$. Similarly, its value is determined from the oscillator constant via the root-mean-square charge  radius of the nucleus with three constituent nucleons, which is expressed as ~\cite{Song:2012cd}
\begin{widetext}
\begin{eqnarray}\label{r3}
&&\langle r_3^2 \rangle=\frac{1}{2}\frac{ |Q_1 m_1^2(m_2+m_3)+ Q_2m_2^2(m_3+m_1)+ Q_3 m_3^2(m_1+m_2) |}{\omega(m_1+m_2+m_3)m_1m_2m_3},
\end{eqnarray}
\end{widetext}
 where $Q_1$, $Q_2$ and $Q_3$ are the charges of the three nucleons. 

For the production of triton, we  consider the two  production  channels of  $p+n+n\rightarrow t$ (3-body process) and  $d+p\rightarrow t$ (2-body process). Here the deuteron in the latter process is treated as a point-like particle with its phase-space distribution given by that obtained from the coalescence of protons and neutrons. Note that the final triton yield is the summation over the 2-body and 3-body processes under the assumption that   the  coalescence processes occur instantaneously and monodirectionally~\cite{Chen:2003ava,Chen:2003qj}.  Alternatively, if one assumes  that  the  triton yield in the coalescence model using nucleons from a thermally and chemically equilibrated emission source is the same as in the statistical model with the triton binding energy neglected, then the two coalescence processes $p+n+n\rightarrow t$ and $d+p\rightarrow t$ would give the same triton yield~\cite{Sun:2017xrx}. In this case,  only one of the two processes   should be considered in the coalescence model.  In this work, we will quantify   triton production from the 2-body and 3-body   processes separately. Because of the very small number of (anti-)deuterons and tritons produced in heavy ion collisions, the  protons  and anti-protons participating  in the coalescence processes have negligible effects in calculating  the final (anti-)proton spectra.

TABLE \ref{tab} provides the statistical factors and the values of the width parameters in the Wigner functions for deuterons and tritons as well as the empirical values of their charge radii and the resulting oscillator constants. 

\begin{table}[h]
\caption{ Statistical factor ($g$),  charge  radius ($R$), oscillator frequency ($\omega$) and width parameter  ($\sigma_\rho$, $\sigma_\lambda$) for  (anti-)deuteron and  (anti-)triton.  Charge radii  are taken from Ref.~\cite{Angeli:2013epw}. }
\label{tab}
\begin{tabular}{c|cccccc}
\hline\hline
Nucleus & $g$ & R (fm) & $\omega$ (sec$^{-1})$ & $\sigma_\rho, \sigma_\lambda$ (fm) \\
\hline
$p+n\rightarrow$deuteron & 3/4 & 2.1421 & 0.1739 & 2.473 \\
$p+n+n\rightarrow$triton & 1/4 & 1.7591 & 0.3438 & 1.759 \\
$d+n\rightarrow$ tritin & 1/3 & 1.7591 & 0.2149 & 1.927 \\
\hline\hline
\end{tabular}
\end{table}

%\subsection{The {\tt iEBE-MUSIC} hybrid framework for collision dynamics and particle production}

\subsection{The {\tt iEBE-MUSIC} hybrid model for collision dynamics and particle production}

\label{sec:music}

For the phase-space distributions of (anti-)protons and (anti-)neutrons used in the  coalescence model calculations of light (anti-)nuclei production at RHIC BES energies, we employ the {\tt iEBE-MUSIC} hybrid  model~\cite{iEBEMUSIC} to describe the collision dynamics until the  kinetic freeze-out.  {\tt iEBE-MUSIC} is a generic event generator to simulate the QGP collective dynamics and soft hadrons production in relativistic heavy-ion collisions. At the RHIC BES energies, this  hybrid model   uses  a 3D Monte-Carlo (MC) Glauber initial condition to  dynamically  deposit  energy, momentum, and net baryon densities into the evolving fluid system as the two colliding nuclei are penetrating through each other~\cite{Shen:2017bsr, 3DGlauber}.
The collective expansion of the QGP fireball and the evolution of the conserved net-baryon current are simulated  by a (3+1)D viscous hydrodynamic model {\tt MUSIC}~\cite{Schenke:2010rr, Schenke:2010nt, Paquet:2015lta, Denicol:2018wdp, MUSIC}.  As the QGP expands and  transitions to  the dilute hadronic phase, the fluid dynamic description is switched to a microscopic hadron cascade model, {\tt UrQMD}~\cite{Bass:1998ca, Bleicher:1999xi, UrQMD}, to simulate the succeeding  evolution and decoupling of the hadronic matter.

More specifically, the dynamical initial condition is simulated by the 3D Monte-Carlo Glauber model  on  an event-by-event basis~\cite{Shen:2017bsr}, and the space-time and momentum distributions of the initial energy-momentum tensor and  net baryon  charge current are provided by the classical string deceleration model~\cite{Bialas:2016epd,Shen:2017bsr}.  In order to reproduce the  pseudo-rapidity distributions of charged hadrons for  Au+Au Collisions at $\sqrt{s_\mathrm{NN}}= 7.7-200$ GeV, we use the parameterized rapidity loss function given in Ref.~\cite{Shen:2018pty}. We further introduce additional baryon charge fluctuations according to the string junction model~\cite{Kharzeev:1996sq, Shenfuture1}, which helps to achieve a good description of the measured rapidity distributions of net protons. The detailed implementation of  this initial condition model and systematic phenomenological impacts will be reported in an upcoming work~\cite{Shenfuture1}.

With such dynamical initial conditions, the hydrodynamic   equations for  the evolution of the energy-momentum tensor and the net baryon current   are then solved with the inclusion of  source terms~\cite{Shen:2017bsr}. Here we use a the crossover equation of state ({\tt NEOS-BQS}) for the QCD matter at finite chemical potentials, potentials that is constructed from recent lattice~\cite{Borsanyi:2011sw, Borsanyi:2013bia, Ding:2015fca, Bazavov:2017dus, Monnai:2019hkn}. This EoS  is obtained by imposing  the strangeness neutrality condition  of vanishing net strangeness density,  $n_s=0$, and setting the net electric charge-to-baryon density ratio to $n_Q/n_B=0.4$~\cite{Monnai:2019hkn}. Note that  this  EoS does not contain a QCD critical point since the model calculations in this paper aim to provide clean baseline results  without any  effects from  critical fluctuations for the related measurements of light nuclei at  the RHIC BES program. We leave the study of the influence of a critical point or critical fluctuations to  future  works. Following Refs.~\cite{Shen:2018pty, Monnai:2019hkn}, we only consider the shear viscous effects in the hydrodynamic evolution with the specific shear viscosity set to a constant value $\frac{\eta T}{e+P}=0.08$. The shear stress tensor is evolved according to a set of    relaxation type of  equations up to the second order in spatial gradients~\cite{Denicol:2012cn, Denicol:2018wdp}. For simplicity, the effects from bulk viscosity and charge diffusion are neglected in this work.

In {\tt iEBE-MUSIC}, the  Cooper-Frye  particlization of the fluid cells is performed  on a hyper-surface with a constant energy density of $e_{sw}=$0.26 GeV/fm$^3$ using the open-source code package {\tt iSS} \cite{Shen:2014vra, iSS_code}.  The produced hadrons are then fed into the hadron cascade model, {\tt UrQMD}, for further scatterings and decays until  their  kinetic freeze-outs.  Finally,  we  obtain the  freeze-out  phase-space distributions of nucleons for the  coalescence model calculations.

\begin{figure*}[pht]
  \centering \includegraphics[scale=0.75]{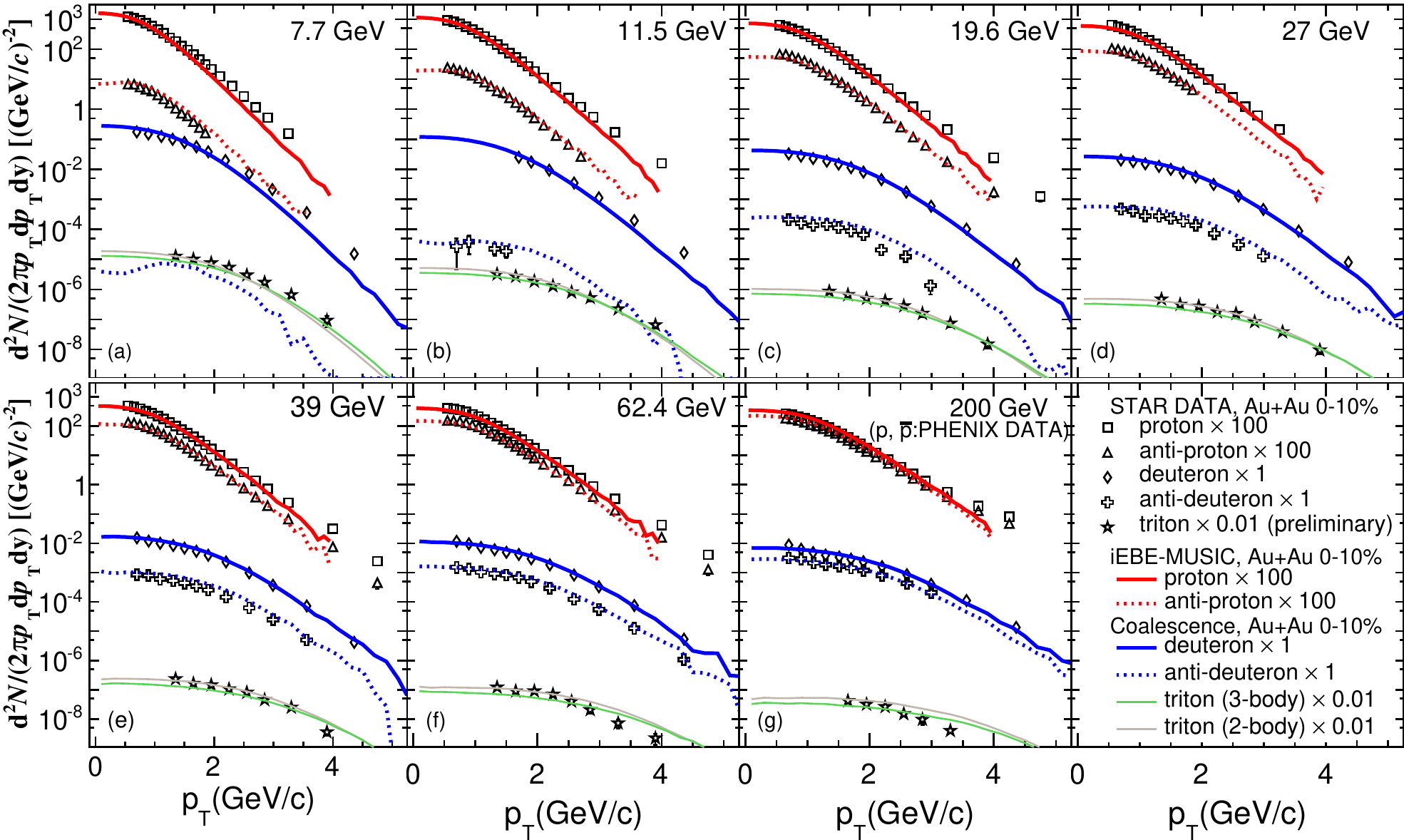}
  \caption{(Color online) Transverse momentum spectra of  (anti-)protons,  (anti-)deuterons and tritons in 0-10\% Au + Au collisions at $\sqrt{s_\mathrm{NN}}=$ 7.7, 11.5, 19.6, 27, 39, 62.4, and 200 GeV. The data for (anti-)deuterons  and tritons are taken from the STAR  Collaboration~\cite{Adamczyk:2017nof,Adam:2019wnb,Zhang:2019wun}, and the data for (anti-)protons are taken from the STAR and PHENIX Collaborations~\cite{Adler:2003cb}. }
  \label{fig:sepctratdp}
\end{figure*}

A quantitative coalescence model calculation for   light nuclei production requires realistic  phase-space distributions of nucleons at the kinetic freeze-out~\cite{Zhao:2017yhj,Zhao:2018lyf}.  Therefore, it is necessary to achieve a    good  description of the identified particle $\langle p_T \rangle$ and $p_T$-spectra. Here, we emphasize that the {\tt iEBE-MUSIC}  hybrid model  employed in this paper can capture both the longitudinal and transverse dynamics of the collision system.  This  hybrid model has achieved a consistent description of soft particle production in the most central Au-Au collisions  at $\sqrt{s_\mathrm{NN}}= 7.7-200$ GeV,  as demonstrated in Ref.~\cite{Shen:2018pty, Shen:2020gef} and   Fig.~\ref{fig:auauspectra} in the Appendix.  The description of various  flow observables within the {\tt iEBE-MUSIC}  hybrid model  will be reported in the upcoming works~\cite{Shenfuture1}.

\section{RESULTS}\label{sec:results}

\begin{figure}[t]
  \includegraphics[scale=0.4]{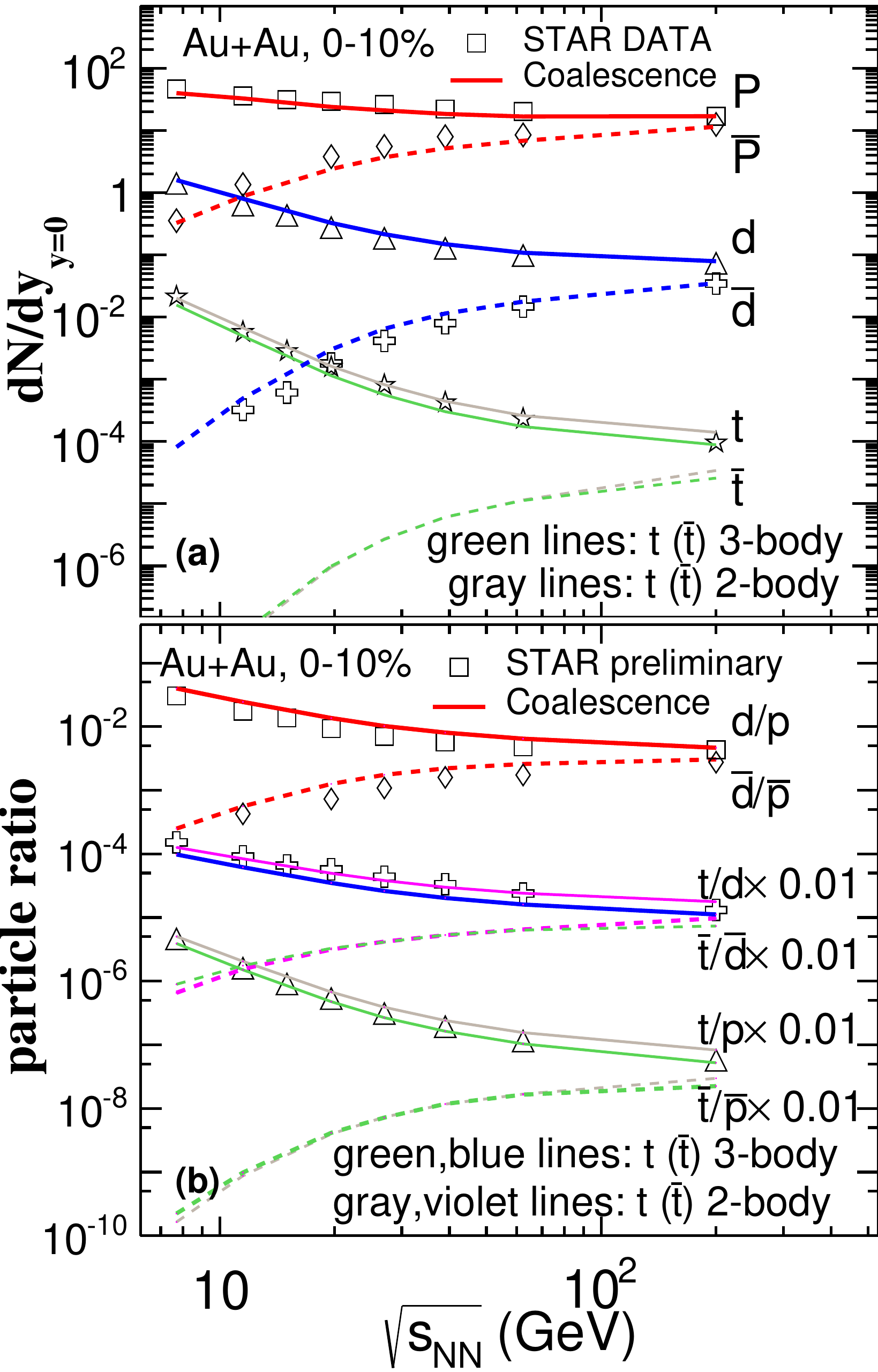}
  \caption{(Color online) Collision energy dependence of (a) $dN/dy$ of (anti-)protons,  (anti-)deuterons, and   (anti-)tritons  at mid-rapidity and (b) the particle ratio $d/p$, $\bar{d}/\bar{p}$, $t/d$, $\bar{t}/\bar{d}$,  $t/p$ and $\bar{t}/\bar{p}$  in  0-10\% Au + Au collisions.
The experimental data for (anti-)protons, (anti-)deuterons, and tritons are taken from~\cite{Adamczyk:2017nof,Adler:2003cb,Adam:2019wnb,Zhang:2019wun}. }
  \label{fig:dndy}
\end{figure}

In this section, we  study  the  transverse momentum spectra  and  particle yield $dN/dy$ at mid-rapidity, coalescence parameters $\sqrt[A-1]{B_{A}}$ $(A=2,3)$ and yield ratios of light (anti-)nuclei in 0-10\%  Au+Au collisions at $\sqrt{s_\mathrm{NN}}=$7.7, 11.5, 19.6, 27, 39, 62.4, and 200 GeV. Simulation results are  calculated from the coalescence model using the phase-space distributions of (anti-)protons and (anti-)neutrons generated from  the  {\tt iEBE-MUSIC}  hybrid model.

\subsection{Transverse Momentum Spectra and $dN/dy$}

Figure~\ref{fig:sepctratdp} shows the transverse momentum spectra of (anti-)protons, (anti-)deuterons and tritons  in the most central (0-10\%)\footnote{Here, we cut the centrality bins in {{\tt iEBE-MUSIC}} calculations according to the impact parameter $\rm b$ in the initial state. The upper limits  for $\rm b$ are 4.60 fm, 4.61 fm, 4.62 fm, 4.65 fm, 4.67 fm, 4.68 fm, and 4.70 fm for 0-10\% Au+Au collisions at $\sqrt{s_\mathrm{NN}}=$ 7.7, 11.5, 19.6, 27, 39, 62.4, and 200 GeV, respectively.
 } Au + Au collisions at $\sqrt{s_\mathrm{NN}}=$ 7.7, 11.5, 19.6, 27, 39, 62.4, and 200 GeV.
For the spectra of protons  and anti-protons\footnote{The data  on  (anti-)protons for Au+Au collisions at $\sqrt{s_\mathrm{NN}}= 7.7-62.4$ GeV  are taken from the STAR measurements~\cite{Adamczyk:2017nof} and   those  at $\sqrt{s_\mathrm{NN}}=$ 200 GeV are taken from the PHENIX measurements~\cite{Adler:2003cb}. All data have been corrected by subtracting the feed-down contributions from hyperon weak decays.}, the  {{\tt iEBE-MUSIC}} hybrid  model  gives a  quantitative description of the measured data below 2.5 GeV, but slightly underestimates the data above 3 GeV. At the high $p_T$ region, the quark recombination  process~\cite{Greco:2003mm,Greco:2003vf,Fries:2003kq,Hwa:2004ng,Fries:2008hs,Zhao:2019ehg}, not included in  the Cooper-Frye particlization, gradually becomes important. With the phase-space distributions of (anti-)protons and (anti-)neutrons at kinetic freeze-out, we calculate the spectra of (anti-)deuterons and tritons using the nucleon coalescence model. As shown with the blue solid and dotted lines in Fig.~\ref{fig:sepctratdp}, our model calculations nicely reproduce the $p_T$-spectra of deuterons and anti-deuterons measured by the STAR Collaboration over a wide range of collision energies. The good theoretical descriptions extend to higher $p_T$ at higher collision energies as a result of the stronger  hydrodynamic radial flow.
The transverse momentum spectra of tritons are calculated using  both  the $p+n+n\rightarrow t$ (3-body)  and the $d+p\rightarrow t$ (2-body) coalescence processes.  Our results  from the 3-body process  reasonably describe the STAR data in Au+Au collisions at $\sqrt{s_\mathrm{NN}}= 7.7-39$ GeV. Including the additional 2-body channel would overestimate the triton yield by a factor of 2. Hence our calculations indicate that  triton yield  at RHIC BES is  close to the  thermal equilibrium, consistent with the expectation from the statistical model~\cite{Andronic:2017pug,Sun:2020pjz}. At $\sqrt{s_\mathrm{NN}}=$ 62.4 and 200 GeV, the slopes of  the calculated  triton $p_T$-spectra are slightly harder than those of the measured  ones, which might be caused by the stronger radial flow at $\sqrt{s_\mathrm{NN}}=$ 62.4 and 200 GeV in our model calculations.

Figure~\ref{fig:dndy}a shows the dependence of the mid-rapidity particle yields for (anti-)protons, (anti-)deuterons, and (anti-)tritons on collision energy. Our simulations quantitatively reproduce the STAR measurements within 10\%. The final proton yields are larger at lower collision energies because of the interplay between the effects of baryon charge transport and  the thermal production of nucleons. The 3D MC-Glauber model, with the dynamical initialization scheme and string junction fluctuations  for net baryon charges  gives a realistic estimation of initial baryon stopping. For the proton yields at lower collision energies, the contributions from the initial baryon stopping and baryon current evolution during the hydrodynamic phase gradually overwhelms those from the thermal production at particlization.  The calculated $dN/dy$ of deuterons, (anti-)deuterons, tritons, and (anti-)tritons also show a similar  dependence on the  collision  energy, which again gives a reasonable description of the STAR data.
%Besides, the yields of tritons generated from $p+n+n\rightarrow t$ (3-body process) are about 20\% larger than those from $d+p\rightarrow t$ (2-body process).

Figure~\ref{fig:dndy}b shows the energy dependence of the yield  ratios of light (anti-)nuclei to (anti-)protons. In general, these calculated ratios agree with the measured  data in the most central Au+Au collisions at $\sqrt{s_\mathrm{NN}}= 7.7-200$ GeV  within a 20\% accuracy. Our calculation overestimates the $d/p$ and $\bar{d}/\bar{p}$ ratios by 15\% and 20\%, respectively. The coalescence model nicely reproduces the $t/p$ ratios with tritons produced from the 3-body process, while underestimates the $t/d$ ratios by 10\%.
%while the 2-body production channel overestimates the ratios by 15\%.
%In the meantime, the coalescence model underestimates the $t/d$ ratios by 10\% with tritons produced from the 3-body process. but well describes the $t/d$ ratios with tritons produced from the 2-body process.

\subsection{Coalescence Parameters and Light Nuclei Yield Ratios}

In the coalescence picture, the invariant yield of light nuclei with the mass number $A=Z+N$ is proportional to the invariant yields of protons and neutrons according to
\begin{eqnarray}
E_{A}\frac{\mathrm{d}^{3}N_{A}}{\mathrm{d}p_{A}^{3}}&=&B_{A}{\left(E_{\mathrm{p}}\frac{\mathrm{d}^{3}N_{\mathrm{p}}}{\mathrm{d}p_{\mathrm{p}}^{3}}\right)^{Z}}{\left(E_{\mathrm{n}}\frac{\mathrm{d}^{3}N_{\mathrm{n}}}{\mathrm{d}p_{\mathrm{n}}^{3}}\right)^{A-Z}}\nonumber\\
&\approx&B_{A}{\left(E_{\mathrm{p}}\frac{\mathrm{d}^{3}N_{\mathrm{p}}}{\mathrm{d}p_{\mathrm{p}}^{3}}\right)^{A}}\left\vert_{\vec{p}_{\mathrm{p}}=\vec{p}_{\mathrm{n}}=\frac{\vec{p}_{A}}{A}} \right. ,
\label{eq:BA}
\end{eqnarray}
where $\vec{p}_{\mathrm{p,n}}$ are  the proton and neutron  momenta and $E_{p,n}$ are their energies. The coalescence parameter $B_A$ characterizes the coalescence probability and  is related to the effective volume, $V_{\text{eff}}$, of the hadronic emission  source~\cite{Csernai:1986qf,Scheibl:1998tk,Bellini:2018epz},
\begin{equation}\label{b2vsv}
{B_A}\propto V_{\text{eff}}^{1-A}.
\end{equation}

\begin{figure}[b]
  \includegraphics[scale=0.35]{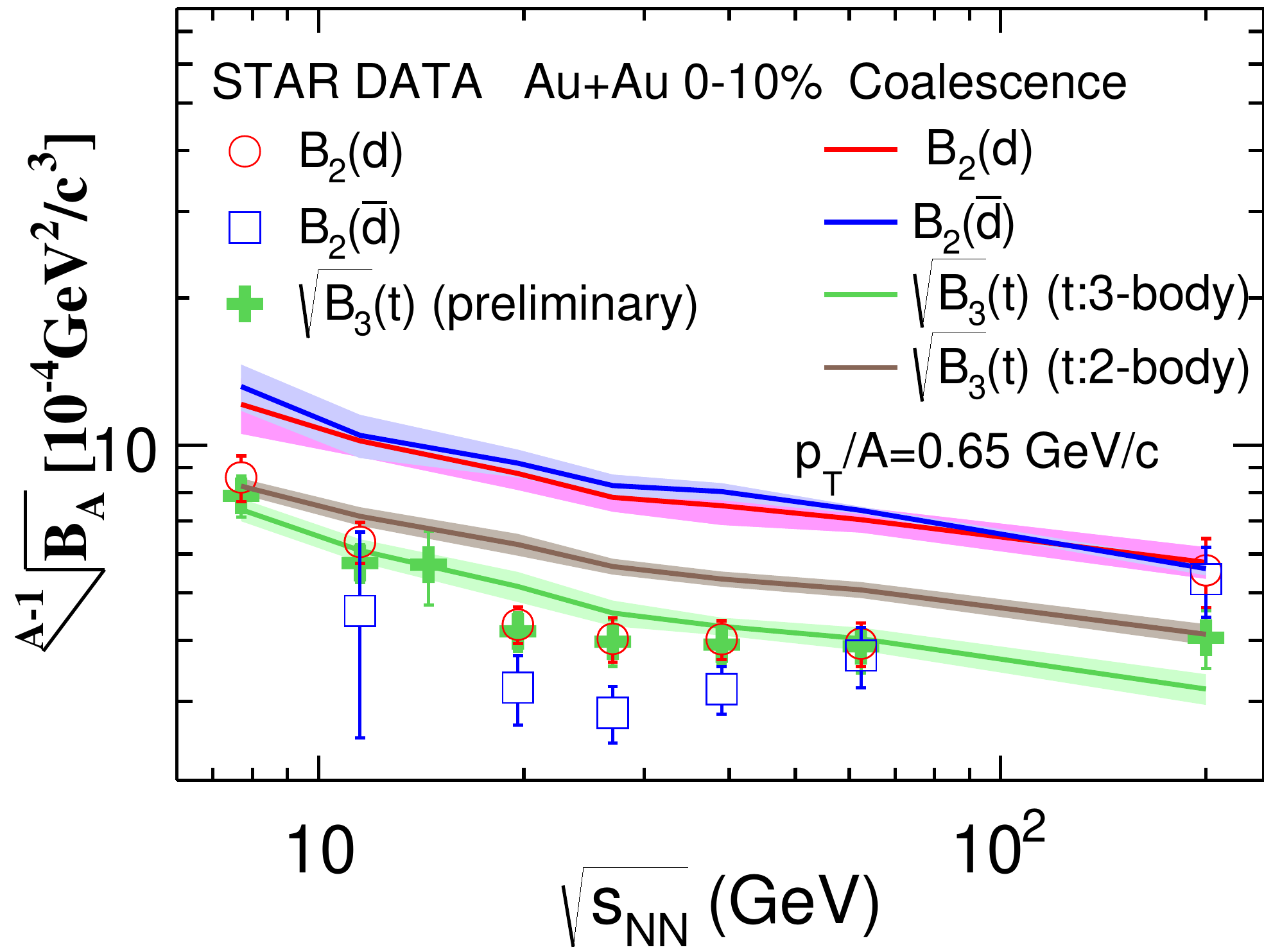}
  \caption{(Color online) Collision energy dependence of the coalescence parameters $B_2(d)$, $B_2(\bar{d})$ and $\sqrt{B_3(t)}$ at $p_T/A=$ 0.65 GeV in 0-10\%  Au+Au collisions, calculated by  the  coalescence model. The values extracted from the experimental  data  for $B_2(d)$,   $B_2(\bar{d})$ and  $B_3(t)$  are taken from~\cite{Adam:2019wnb,Zhang:2019wun}. }
  \label{fig:bn}
\end{figure}

Figure~\ref{fig:bn} shows  the collision energy dependence of the coalescence parameters $B_2(d)$, $B_2(\bar{d})$ and $\sqrt{B_3(t)}$ at $p_T/A$=0.65 GeV in the most central Au+Au collisions, with $A=2$ for (anti-)deuterons and $A=3$ for tritons. The measured $B_2(d)$ and $B_2(\bar{d})$ from the STAR collaboration~\cite{Adam:2019wnb} show a non-monotonic dependence on the collision energy with a dip located around $\sqrt{s_\mathrm{NN}}= 20-40$ GeV, which might indicate a dramatic change of the equation of state  in the produced matter  at these collision energies~\cite{Adam:2019wnb, Lacey:2014wqa}. In contrast, our coalescence model calculations, using the phase-phase distributions of protons and neutrons generated from  the {\tt iEBE-MUSIC} hybrid model   with a crossover  EoS  in the hydrodynamics, gives a monotonically decreasing $B_2(d)$ and $B_2(\bar{d})$, which is because the overall sizes of the emitting source of nucleons increases monotonically with the collision energy in our model.    Also, our  model overestimates the values $B_2(d)$ and $B_2(\bar{d})$ by $\sim$50\% for $\sqrt{s_\mathrm{NN}}= 20-62.4$ GeV, because our calculations overestimate the yield of deuterons by 10\% while underestimate the proton yield by 15\%. The relative ratios between proton and deuteron yields are sensitive to the phase-space distribution of nucleons at the kinetic freeze-out. Therefore, the experimental measurements of $B_2(d)$ and $B_2(\bar{d})$ can set strong constraints on the spatial-momentum correlations of nucleons in the hadronic phase. In addition, the measured $B_2(d)$ and  $B_2(\bar{d})$ curves  as functions of the collision energy  show  a clear separation, while these curves from our calculations almost overlap. We  note  that  these coalescence parameters have recently also been studied in Ref.~\cite{Oliinychenko:2020znl}  by using the hydrodynamics + SMASH hadronic transport model with event-averaged 3D initial conditions based on collision geometry \cite{Shen:2020jwv}.  Instead of production from nucleon coalescence, (anti-)deuterons in this study are treated as dynamic degrees of freedom through the pion catalysis reactions $\pi d \leftrightarrow \pi pN$ with large cross sections. The resulting (anti-)deuteron yields and spectra were studied for the most central Au+Au collisions at $\sqrt{s_\mathrm{NN}}=7-200$ GeV.  It is pointed out in this study  that the weak decay corrections to the proton spectrum need careful attention as they  could potentially lead to a minimum  in the collision energy dependence of the coalescence parameters $B_2(d)$ and $B_2(\bar{d})$.

As expected from Eq.~(\ref{b2vsv}), the calculated $\sqrt{B_3(t)}$ curve  shows a similar trend as the $B_2(d)$ curve, which monotonically increases with the decrease of the collision energy. This is also consistent with the calculated flat yield ratio $N_tN_p/N_d^2$ described below. Note that our coalescence model calculations with the 3-body process roughly describe the magnitude of the measured $\sqrt{B_3(t)}$ curve in the most central Au+ Au collisions at $\sqrt{s_\mathrm{NN}}= 7.7-200$ GeV, but not the non-monotonic behavior.
In our calculations, the values of $B_2(\bar{d}/d)$ are larger than  those  of $\sqrt{B_3(t)}$, which suggests that  tritons and deuterons are produced with  different degrees of sensitivity to the nucleon phase-space distributions in the coalescence process. Including triton production from the 2-body  coalescence process   increasesthe values of $\sqrt{B_3(t)}$, which makes them closer to the  values of $B_2(\bar{d}/d)$ in our calculations.

\begin{figure}[t]
  \includegraphics[scale=0.35]{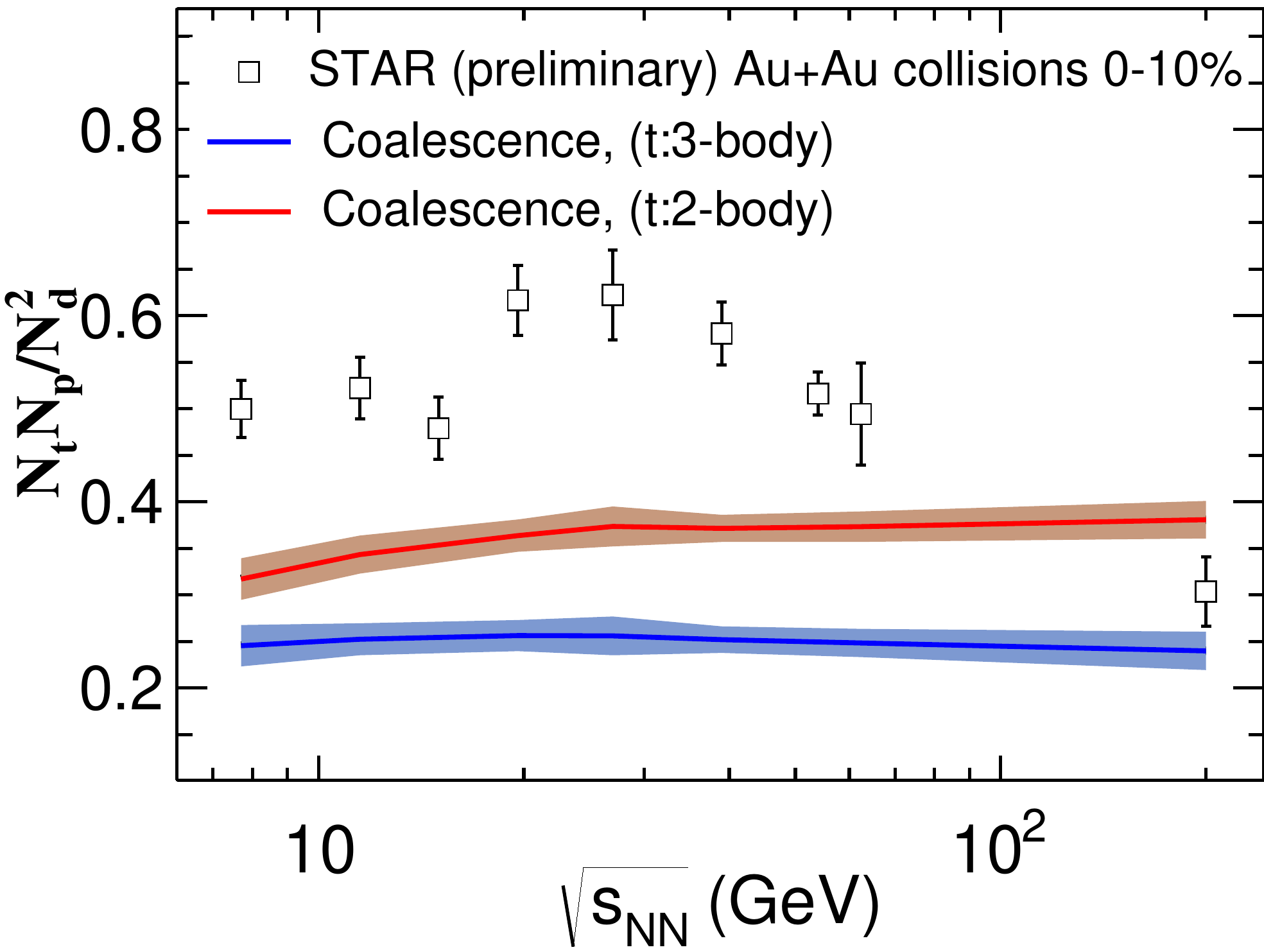}
  \caption   {(Color online) Collision energy dependence of the yield ratio $N_t  N_p/(N^2_d)$ in 0-10\% Au+Au collisions  calculated  from  the  coalescence model. The data is taken from Ref.~\cite{Zhang:2019wun}. }
  \label{fig:ratio}
\end{figure}

Recently, the yield ratio of light nuclei, $N_t  N_p/N_d^2$, in heavy ion collisions has been  suggested  as a sensitive  probe to the neutron density fluctuation   associated with the first-order QGP to hadronic matter  phase transition and the possible critical point of the hot  and baryon-rich  QCD matter~\cite{Sun:2017xrx,Sun:2018jhg,Shuryak:2018lgd,Shuryak:2019ikv}. Figure~\ref{fig:ratio} shows the $N_t  N_p/N_d^2$ ratio as a function of  the  collision   energy from the experiments  by the STAR  Collaboration and  from  our coalescence model  calculations. The measured $N_t  N_p/N_d^2$  ratio shows a non-monotonic behavior with a peak located around $\sqrt{s_\mathrm{NN}}=$ 20 GeV~\cite{Zhang:2019wun}, which might indicate  a non-trivial collision energy dependence of the baryon density fluctuations~\cite{Sun:2017xrx,Sun:2018jhg}. In contrast, the calculated $N_t  N_p/N_d^2$ ratios for both cases of  2-body and 3-body coalescence processes are  almost flat  in their  collision energy  dependence, and this is due to the absence of any non-trivial baryon density fluctuations associated with the critical point as a result of using a  crossover  type EoS in the {{\tt iEBE-MUSIC}} hybrid model.  As to the  yield ratio  $N_t  N_p/N_d^2$, the two-body process slightly overestimates  whereas the  three-body process  slightly underestimates the measured value at 200 GeV.  Both processes greatly underestimate, however, the measured value at  $\sqrt{s_\mathrm{NN}} \leq 62.4$ GeV.

Figure~\ref{fig:ratio} further shows that the yield ratio  $N_t  N_p/N_d^2$ with tritons produced from the 2-body process is larger than that with tritons produced from the 3-body process  in our model, which is a consequence of the non-trivial spatial-momentum correlations in the nucleon phase-space distributions from our  {{\tt iEBE-MUSIC}} hybrid model.  It is shown in  Ref.~\cite{Sun:2018mqq}   that the  yield  ratios from these two   processes  would be the same if the nucleon phase-space distributions are uniform in the coordinate space.  We emphasis that our model does not contain any effects from a critical point, which thus provides  the  non-critical baseline results for  the yields of these light nuclei  in heavy ion collisions  at the RHIC BES energies. For a better explanation  of  the observed non-monotonic behavior of  $N_t  N_p/N_d^2$, $B_2(d)$,  $B_2(\bar{d})$ and $\sqrt{B_3}(t)$  in their collision energy dependence, a dynamical model with critical fluctuations or the effects of critical point is required.
%In principle, once the phase space distribution of nucleon is uniform and the momentum space and coordinate space are factorized, then the 2-body channel and 3-body channel contributions to  yield of triton would be the same~\cite{Sun:2018mqq}.  However, in our model simulations  the phase-space distributions of nucleon at kinetic freeze-out fluctuate event-by-event and the collective motion of the bulk medium  introduces the non-trivial correlations between the distributions of momentum space and coordinate space of nucleons. Such dynamical effects might cause the yield of tritons from 2-body channel larger than that from 3-body channel here. Nevertheless, our calculation with either of these two processes underestimates the data at low collision energies, but  close to the measurement at $\sqrt{s_\mathrm{NN}}=$ 200 GeV.

We note that our result on the yield ratio $N_tN_p/N_d^2$ is similar to those found in Ref.~\cite{Liu:2019nii}, which is based on a simple phase-space coalescence model using nucleons from the JAM hadronic cascade model~\cite{Nara:1999dz} and in Ref.~\cite{Sun:2020uoj}, which is based on a coalescence model similar to that in the present study with nucleons from a multiphase transport (AMPT) model~\cite{Lin:2004en}. 

Although a non-monotonic collision energy dependence of the yield ratio $N_tN_p/N_d^2$  has been  reported in Ref.~\cite{Deng:2020zxo} from a coalescence model study using nucleons from the UrQMD model~\cite{Bass:1998ca}, the result is puzzling because of the unexpected very different nucleon and light nuclei rapidity distributions predicted from this study.

\section{Summary}\label{sec:summary}

In this paper, we have  used the nucleon coalescence model to study  light nuclei production in the most central Au+Au collisions at $\sqrt{s_\mathrm{NN}}=$ 7.7, 11.5, 19.6, 27, 39, 62.4 and 200 GeV. The input phase-space distributions of (anti-)protons and (anti-)neutrons at kinetic freeze-out for the coalescence calculations are generated from the {{\tt iEBE-MUSIC}} hybrid model using three dimensional dynamical initial conditions and a  crossover   EoS.  These comprehensive  simulations can nicely reproduce the measured $p_T$-spectra of (anti-)pions, (anti-)kaons, and (anti-)protons for Au+Au collisions at $\sqrt{s_\mathrm{NN}}= 7.7-200$ GeV (as shown in the appendix and in Ref.~\cite{Shen:2020gef}).   We have found that the  subsequent  coalescence model calculations can reproduce the  measured  $p_T$-spectra and $dN/dy$ of (anti-)deuterons and (anti-)tritons and the particle ratios of $t/p$  within 10\% of accuracy.  However, the deviations between the calculated and measured  particle ratios of $d/p$, $\bar{d}/\bar{p}$, and $t/d$ increase  to 15\%, 20\%, and 10\%, respectively.

Although the coalescence model reasonably describes the $p_T$-spectra and  yields of light nuclei at various collision energies, the predicted coalescence parameters of  (anti-)deuterons and tritons, $B_2(d), B_2(\bar{d})$ and $\sqrt{B_3(t)}$, decrease monotonically  with increasing collision energy,   and  the yield ratio $N_t  N_p/N_d^2$ stays  almost  constant with  respect to  the  collision energy. All  these theoretical  results  fail to describe the non-monotonic behavior of the corresponding measurements in experiments. We emphasis that the hydrodynamic part of our calculations with a  crossover EoS for all collision energies  does not generate any dynamical density fluctuations, which are  related to  the  critical point and first-order phase transition, for the  subsequent  nucleon coalescence model calculations. According to Refs.~\cite{Sun:2017xrx,Sun:2018jhg}, non-trivial density fluctuations in the produced hot QCD matter are needed to describe  this   non-monotonic behavior. Our model calculations thus  provide the non-critical baseline results for  comparisons with  related light nuclei measurements at the RHIC BES program. We leave the implementation of an EoS with a critical point in the  hydrodynamic evolution and the inclusion of dynamical density fluctuations to  future  studies.

\begin{figure*}[t]
  \centering \includegraphics[scale=0.75]{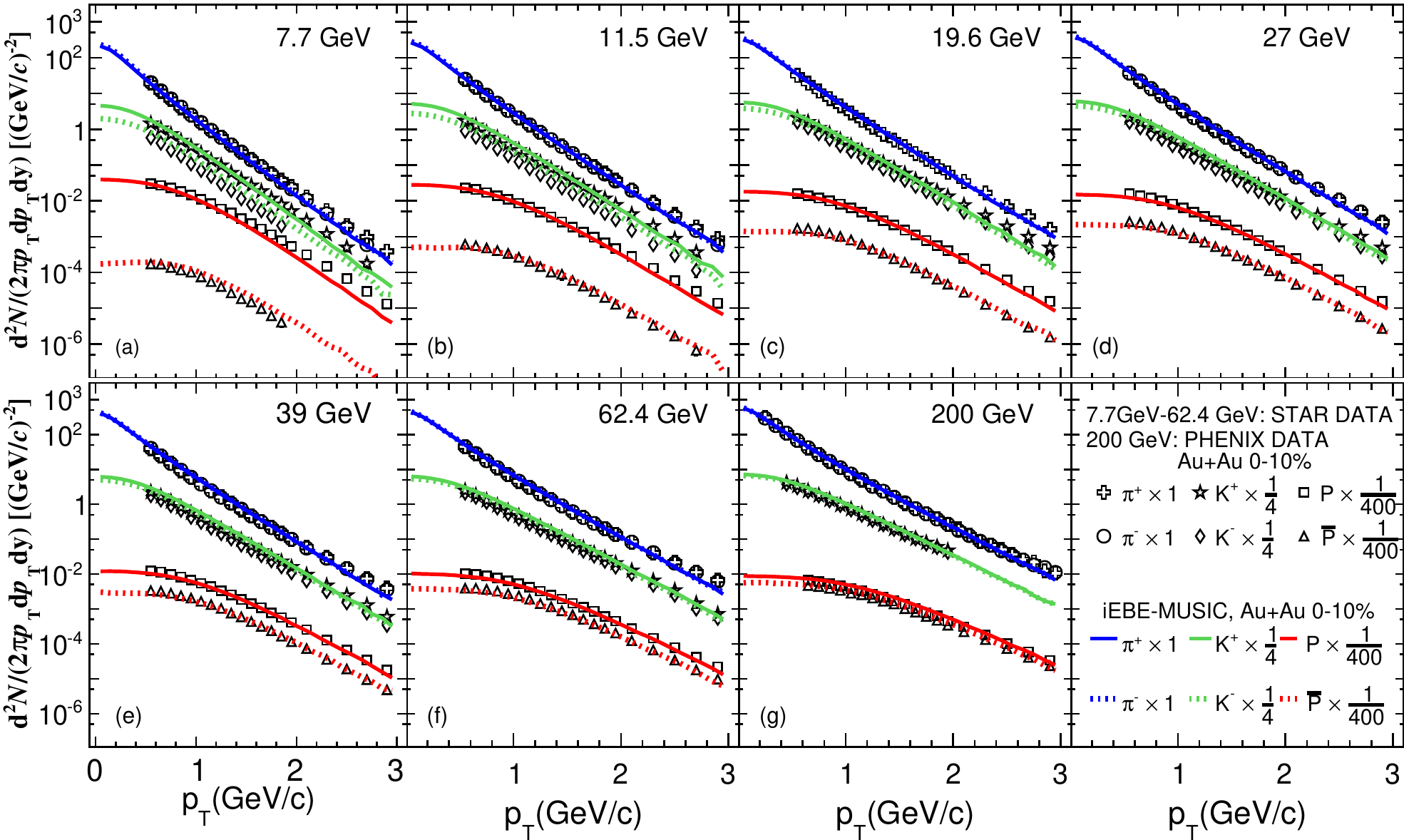}
  \caption{(Color online) Transverse momentum spectra of (anti-)pions, (anti-)kaons, and (anti-)protons in 0-10\% Au + Au collisions at $\sqrt{s_\mathrm{NN}}=$ 7.7, 11.5 19.6, 27, 39, 62.4, and 200 GeV, calculated from the {{\tt iEBE-MUSIC}} hybrid model.  The data are taken from the  STAR and PHENIX Collaborations~\cite{Adamczyk:2017nof, Adler:2003cb}.}
  \label{fig:auauspectra}
\end{figure*}

\section*{Acknowledgements}
We thank  X.~F.~Luo, N.~Yu and D.~W.~Zhang for providing the STAR data  as well as  K.~Murase, D.~Oliinychenko, K.~J.~Sun and S.~Wu for discussions. W.~Z., Q.~L. and H.~S. are supported by the NSFC under grant Nos. 11675004.  C.~S is supported in part by the U.S. Department of Energy (DOE)  under grant number DE-SC0013460 and in part by the National Science Foundation (NSF) under grant number PHY-2012922. C.M. K. is supported by US DOE under Award No. DE-SC0015266 and the Welch Foundation under Grant No. A-1358. This work is also supported in part by the U.S. Department of Energy, Office of Science, Office of Nuclear Physics, within the framework of the Beam Energy Scan Theory (BEST) Topical Collaboration. C.~S further acknowledge computing resources of the National Energy Research Scientific Computing Center, which is supported by the Office of Science of the U.S. Department of Energy under Contract No. DE-AC02-05CH11231. W. Z. and H. S. also gratefully acknowledge the extensive computing resources provided by the Super-computing Center of Chinese Academy of Science (SCCAS), Tianhe-1A from the National Supercomputing Center in Tianjin, China and the High-performance Computing Platform of Peking University.

\section*{Appendix: $p_T$-spectra of (anti-)pions, (anti-)kaons and (anti-)protons}
\label{sec:appendix}

In this appendix, we present the {{\tt iEBE-MUSIC}} hybrid model calculations  using the dynamical initialization and string junction fluctuations for net baryon charges to study  the $p_T$-spectra of (anti-)pions, (anti-)kaons, and (anti-)protons in 0-10\% central Au + Au collisions at $\sqrt{s_\mathrm{NN}}=$ 7.7, 11.5 19.6, 27, 39, 62.4, and 200 GeV. Figure~\ref{fig:auauspectra} shows that  this model  gives a good description of the $p_T$-spectra of these identified hadrons. Such quantitative descriptions, especially for the $p_T$ spectra of protons and anti-protons, demonstrates that this three-dimensional hybrid model, without any critical fluctuations, can provide a reliable phase-space distributions of nucleons   for the  subsequent coalescence model calculations of light nuclei production  at various collision energies in the RHIC BES program.

%\newpage
\bibliography{bibliography}

\begin{thebibliography}{102}
\expandafter\ifx\csname natexlab\endcsname\relax\def\natexlab#1{#1}\fi
\expandafter\ifx\csname bibnamefont\endcsname\relax
  \def\bibnamefont#1{#1}\fi
\expandafter\ifx\csname bibfnamefont\endcsname\relax
  \def\bibfnamefont#1{#1}\fi
\expandafter\ifx\csname citenamefont\endcsname\relax
  \def\citenamefont#1{#1}\fi
\expandafter\ifx\csname url\endcsname\relax
  \def\url#1{\texttt{#1}}\fi
\expandafter\ifx\csname urlprefix\endcsname\relax\def\urlprefix{URL }\fi
\providecommand{\bibinfo}[2]{#2}
\providecommand{\eprint}[2][]{\url{#2}}

\bibitem[{\citenamefont{Stephanov}(2004)}]{Stephanov:2004wx}
\bibinfo{author}{\bibfnamefont{M.~A.} \bibnamefont{Stephanov}},
  \bibinfo{journal}{Prog. Theor. Phys. Suppl.} \textbf{\bibinfo{volume}{153}},
  \bibinfo{pages}{139} (\bibinfo{year}{2004}), \bibinfo{note}{[Int. J. Mod.
  Phys.A20,4387(2005)]}, \eprint{hep-ph/0402115}.

\bibitem[{\citenamefont{Stephanov}(2006)}]{Stephanov:2007fk}
\bibinfo{author}{\bibfnamefont{M.}~\bibnamefont{Stephanov}},
  \bibinfo{journal}{PoS} \textbf{\bibinfo{volume}{LAT2006}},
  \bibinfo{pages}{024} (\bibinfo{year}{2006}), \eprint{hep-lat/0701002}.

\bibitem[{\citenamefont{Aoki et~al.}(2006)\citenamefont{Aoki, Endrodi, Fodor,
  Katz, and Szabo}}]{Aoki:2006we}
\bibinfo{author}{\bibfnamefont{Y.}~\bibnamefont{Aoki}},
  \bibinfo{author}{\bibfnamefont{G.}~\bibnamefont{Endrodi}},
  \bibinfo{author}{\bibfnamefont{Z.}~\bibnamefont{Fodor}},
  \bibinfo{author}{\bibfnamefont{S.~D.} \bibnamefont{Katz}}, \bibnamefont{and}
  \bibinfo{author}{\bibfnamefont{K.~K.} \bibnamefont{Szabo}},
  \bibinfo{journal}{Nature} \textbf{\bibinfo{volume}{443}},
  \bibinfo{pages}{675} (\bibinfo{year}{2006}), \eprint{hep-lat/0611014}.

\bibitem[{\citenamefont{Aggarwal et~al.}(2010)}]{Aggarwal:2010cw}
\bibinfo{author}{\bibfnamefont{M.}~\bibnamefont{Aggarwal}} \bibnamefont{et~al.}
  (\bibinfo{collaboration}{STAR}) (\bibinfo{year}{2010}), \eprint{1007.2613}.

\bibitem[{\citenamefont{Fukushima and Hatsuda}(2011)}]{Fukushima:2010bq}
\bibinfo{author}{\bibfnamefont{K.}~\bibnamefont{Fukushima}} \bibnamefont{and}
  \bibinfo{author}{\bibfnamefont{T.}~\bibnamefont{Hatsuda}},
  \bibinfo{journal}{Rept. Prog. Phys.} \textbf{\bibinfo{volume}{74}},
  \bibinfo{pages}{014001} (\bibinfo{year}{2011}), \eprint{1005.4814}.

\bibitem[{\citenamefont{Akiba et~al.}(2015)}]{Akiba:2015jwa}
\bibinfo{author}{\bibfnamefont{Y.}~\bibnamefont{Akiba}} \bibnamefont{et~al.}
  (\bibinfo{year}{2015}), \eprint{1502.02730}.

\bibitem[{\citenamefont{Heinz et~al.}(2015)}]{Heinz:2015tua}
\bibinfo{author}{\bibfnamefont{U.}~\bibnamefont{Heinz}} \bibnamefont{et~al.}
  (\bibinfo{year}{2015}), \eprint{1501.06477}.

\bibitem[{\citenamefont{Asakawa and Kitazawa}(2016)}]{Asakawa:2015ybt}
\bibinfo{author}{\bibfnamefont{M.}~\bibnamefont{Asakawa}} \bibnamefont{and}
  \bibinfo{author}{\bibfnamefont{M.}~\bibnamefont{Kitazawa}},
  \bibinfo{journal}{Prog. Part. Nucl. Phys.} \textbf{\bibinfo{volume}{90}},
  \bibinfo{pages}{299} (\bibinfo{year}{2016}), \eprint{1512.05038}.

\bibitem[{\citenamefont{Luo and Xu}(2017)}]{Luo:2017faz}
\bibinfo{author}{\bibfnamefont{X.}~\bibnamefont{Luo}} \bibnamefont{and}
  \bibinfo{author}{\bibfnamefont{N.}~\bibnamefont{Xu}}, \bibinfo{journal}{Nucl.
  Sci. Tech.} \textbf{\bibinfo{volume}{28}}, \bibinfo{pages}{112}
  (\bibinfo{year}{2017}), \eprint{1701.02105}.

\bibitem[{\citenamefont{Bzdak et~al.}(2019)\citenamefont{Bzdak, Esumi, Koch,
  Liao, Stephanov, and Xu}}]{Bzdak:2019pkr}
\bibinfo{author}{\bibfnamefont{A.}~\bibnamefont{Bzdak}},
  \bibinfo{author}{\bibfnamefont{S.}~\bibnamefont{Esumi}},
  \bibinfo{author}{\bibfnamefont{V.}~\bibnamefont{Koch}},
  \bibinfo{author}{\bibfnamefont{J.}~\bibnamefont{Liao}},
  \bibinfo{author}{\bibfnamefont{M.}~\bibnamefont{Stephanov}},
  \bibnamefont{and} \bibinfo{author}{\bibfnamefont{N.}~\bibnamefont{Xu}}
  (\bibinfo{year}{2019}), \eprint{1906.00936}.

\bibitem[{\citenamefont{Stephanov}(2009)}]{Stephanov:2008qz}
\bibinfo{author}{\bibfnamefont{M.~A.} \bibnamefont{Stephanov}},
  \bibinfo{journal}{Phys. Rev. Lett.} \textbf{\bibinfo{volume}{102}},
  \bibinfo{pages}{032301} (\bibinfo{year}{2009}), \eprint{0809.3450}.

\bibitem[{\citenamefont{Stephanov}(2011)}]{Stephanov:2011pb}
\bibinfo{author}{\bibfnamefont{M.}~\bibnamefont{Stephanov}},
  \bibinfo{journal}{Phys. Rev. Lett.} \textbf{\bibinfo{volume}{107}},
  \bibinfo{pages}{052301} (\bibinfo{year}{2011}), \eprint{1104.1627}.

\bibitem[{\citenamefont{Athanasiou et~al.}(2010)\citenamefont{Athanasiou,
  Rajagopal, and Stephanov}}]{Athanasiou:2010kw}
\bibinfo{author}{\bibfnamefont{C.}~\bibnamefont{Athanasiou}},
  \bibinfo{author}{\bibfnamefont{K.}~\bibnamefont{Rajagopal}},
  \bibnamefont{and}
  \bibinfo{author}{\bibfnamefont{M.}~\bibnamefont{Stephanov}},
  \bibinfo{journal}{Phys. Rev.} \textbf{\bibinfo{volume}{D82}},
  \bibinfo{pages}{074008} (\bibinfo{year}{2010}), \eprint{1006.4636}.

\bibitem[{\citenamefont{Asakawa et~al.}(2009)\citenamefont{Asakawa, Ejiri, and
  Kitazawa}}]{Asakawa:2009aj}
\bibinfo{author}{\bibfnamefont{M.}~\bibnamefont{Asakawa}},
  \bibinfo{author}{\bibfnamefont{S.}~\bibnamefont{Ejiri}}, \bibnamefont{and}
  \bibinfo{author}{\bibfnamefont{M.}~\bibnamefont{Kitazawa}},
  \bibinfo{journal}{Phys. Rev. Lett.} \textbf{\bibinfo{volume}{103}},
  \bibinfo{pages}{262301} (\bibinfo{year}{2009}), \eprint{0904.2089}.

\bibitem[{\citenamefont{Nahrgang et~al.}(2011)\citenamefont{Nahrgang, Leupold,
  Herold, and Bleicher}}]{Nahrgang:2011mg}
\bibinfo{author}{\bibfnamefont{M.}~\bibnamefont{Nahrgang}},
  \bibinfo{author}{\bibfnamefont{S.}~\bibnamefont{Leupold}},
  \bibinfo{author}{\bibfnamefont{C.}~\bibnamefont{Herold}}, \bibnamefont{and}
  \bibinfo{author}{\bibfnamefont{M.}~\bibnamefont{Bleicher}},
  \bibinfo{journal}{Phys. Rev. C} \textbf{\bibinfo{volume}{84}},
  \bibinfo{pages}{024912} (\bibinfo{year}{2011}), \eprint{1105.0622}.

\bibitem[{\citenamefont{Ling and Stephanov}(2016)}]{Ling:2015yau}
\bibinfo{author}{\bibfnamefont{B.}~\bibnamefont{Ling}} \bibnamefont{and}
  \bibinfo{author}{\bibfnamefont{M.~A.} \bibnamefont{Stephanov}},
  \bibinfo{journal}{Phys. Rev. C} \textbf{\bibinfo{volume}{93}},
  \bibinfo{pages}{034915} (\bibinfo{year}{2016}), \eprint{1512.09125}.

\bibitem[{\citenamefont{Jiang et~al.}(2016)\citenamefont{Jiang, Li, and
  Song}}]{Jiang:2015hri}
\bibinfo{author}{\bibfnamefont{L.}~\bibnamefont{Jiang}},
  \bibinfo{author}{\bibfnamefont{P.}~\bibnamefont{Li}}, \bibnamefont{and}
  \bibinfo{author}{\bibfnamefont{H.}~\bibnamefont{Song}},
  \bibinfo{journal}{Phys. Rev. C} \textbf{\bibinfo{volume}{94}},
  \bibinfo{pages}{024918} (\bibinfo{year}{2016}), \eprint{1512.06164}.

\bibitem[{\citenamefont{Jiang et~al.}(2017)\citenamefont{Jiang, Wu, and
  Song}}]{Jiang:2017mji}
\bibinfo{author}{\bibfnamefont{L.}~\bibnamefont{Jiang}},
  \bibinfo{author}{\bibfnamefont{S.}~\bibnamefont{Wu}}, \bibnamefont{and}
  \bibinfo{author}{\bibfnamefont{H.}~\bibnamefont{Song}},
  \bibinfo{journal}{Nucl. Phys. A} \textbf{\bibinfo{volume}{967}},
  \bibinfo{pages}{441} (\bibinfo{year}{2017}), \eprint{1704.04765}.

\bibitem[{\citenamefont{Mukherjee et~al.}(2015)\citenamefont{Mukherjee,
  Venugopalan, and Yin}}]{Mukherjee:2015swa}
\bibinfo{author}{\bibfnamefont{S.}~\bibnamefont{Mukherjee}},
  \bibinfo{author}{\bibfnamefont{R.}~\bibnamefont{Venugopalan}},
  \bibnamefont{and} \bibinfo{author}{\bibfnamefont{Y.}~\bibnamefont{Yin}},
  \bibinfo{journal}{Phys. Rev. C} \textbf{\bibinfo{volume}{92}},
  \bibinfo{pages}{034912} (\bibinfo{year}{2015}), \eprint{1506.00645}.

\bibitem[{\citenamefont{Mukherjee et~al.}(2016)\citenamefont{Mukherjee,
  Venugopalan, and Yin}}]{Mukherjee:2016kyu}
\bibinfo{author}{\bibfnamefont{S.}~\bibnamefont{Mukherjee}},
  \bibinfo{author}{\bibfnamefont{R.}~\bibnamefont{Venugopalan}},
  \bibnamefont{and} \bibinfo{author}{\bibfnamefont{Y.}~\bibnamefont{Yin}},
  \bibinfo{journal}{Phys. Rev. Lett.} \textbf{\bibinfo{volume}{117}},
  \bibinfo{pages}{222301} (\bibinfo{year}{2016}), \eprint{1605.09341}.

\bibitem[{\citenamefont{Stephanov and Yin}(2018)}]{Stephanov:2017ghc}
\bibinfo{author}{\bibfnamefont{M.}~\bibnamefont{Stephanov}} \bibnamefont{and}
  \bibinfo{author}{\bibfnamefont{Y.}~\bibnamefont{Yin}},
  \bibinfo{journal}{Phys. Rev. D} \textbf{\bibinfo{volume}{98}},
  \bibinfo{pages}{036006} (\bibinfo{year}{2018}), \eprint{1712.10305}.

\bibitem[{\citenamefont{Sakaida et~al.}(2017)\citenamefont{Sakaida, Asakawa,
  Fujii, and Kitazawa}}]{Sakaida:2017rtj}
\bibinfo{author}{\bibfnamefont{M.}~\bibnamefont{Sakaida}},
  \bibinfo{author}{\bibfnamefont{M.}~\bibnamefont{Asakawa}},
  \bibinfo{author}{\bibfnamefont{H.}~\bibnamefont{Fujii}}, \bibnamefont{and}
  \bibinfo{author}{\bibfnamefont{M.}~\bibnamefont{Kitazawa}},
  \bibinfo{journal}{Phys. Rev. C} \textbf{\bibinfo{volume}{95}},
  \bibinfo{pages}{064905} (\bibinfo{year}{2017}), \eprint{1703.08008}.

\bibitem[{\citenamefont{Borsanyi et~al.}(2018)\citenamefont{Borsanyi, Fodor,
  Guenther, Katz, Szabo, Pasztor, Portillo, and Ratti}}]{Borsanyi:2018grb}
\bibinfo{author}{\bibfnamefont{S.}~\bibnamefont{Borsanyi}},
  \bibinfo{author}{\bibfnamefont{Z.}~\bibnamefont{Fodor}},
  \bibinfo{author}{\bibfnamefont{J.~N.} \bibnamefont{Guenther}},
  \bibinfo{author}{\bibfnamefont{S.~K.} \bibnamefont{Katz}},
  \bibinfo{author}{\bibfnamefont{K.~K.} \bibnamefont{Szabo}},
  \bibinfo{author}{\bibfnamefont{A.}~\bibnamefont{Pasztor}},
  \bibinfo{author}{\bibfnamefont{I.}~\bibnamefont{Portillo}}, \bibnamefont{and}
  \bibinfo{author}{\bibfnamefont{C.}~\bibnamefont{Ratti}},
  \bibinfo{journal}{JHEP} \textbf{\bibinfo{volume}{10}}, \bibinfo{pages}{205}
  (\bibinfo{year}{2018}), \eprint{1805.04445}.

\bibitem[{\citenamefont{Akamatsu et~al.}(2019)\citenamefont{Akamatsu, Teaney,
  Yan, and Yin}}]{Akamatsu:2018vjr}
\bibinfo{author}{\bibfnamefont{Y.}~\bibnamefont{Akamatsu}},
  \bibinfo{author}{\bibfnamefont{D.}~\bibnamefont{Teaney}},
  \bibinfo{author}{\bibfnamefont{F.}~\bibnamefont{Yan}}, \bibnamefont{and}
  \bibinfo{author}{\bibfnamefont{Y.}~\bibnamefont{Yin}},
  \bibinfo{journal}{Phys. Rev. C} \textbf{\bibinfo{volume}{100}},
  \bibinfo{pages}{044901} (\bibinfo{year}{2019}), \eprint{1811.05081}.

\bibitem[{\citenamefont{Nahrgang et~al.}(2019)\citenamefont{Nahrgang, Bluhm,
  Schaefer, and Bass}}]{Nahrgang:2018afz}
\bibinfo{author}{\bibfnamefont{M.}~\bibnamefont{Nahrgang}},
  \bibinfo{author}{\bibfnamefont{M.}~\bibnamefont{Bluhm}},
  \bibinfo{author}{\bibfnamefont{T.}~\bibnamefont{Schaefer}}, \bibnamefont{and}
  \bibinfo{author}{\bibfnamefont{S.~A.} \bibnamefont{Bass}},
  \bibinfo{journal}{Phys. Rev. D} \textbf{\bibinfo{volume}{99}},
  \bibinfo{pages}{116015} (\bibinfo{year}{2019}), \eprint{1804.05728}.

\bibitem[{\citenamefont{Odyniec}(2019)}]{Odyniec:2019kfh}
\bibinfo{author}{\bibfnamefont{G.}~\bibnamefont{Odyniec}}
  (\bibinfo{collaboration}{STAR}), \bibinfo{journal}{PoS}
  \textbf{\bibinfo{volume}{CORFU2018}}, \bibinfo{pages}{151}
  (\bibinfo{year}{2019}).

\bibitem[{\citenamefont{Luo}(2015)}]{Luo:2015ewa}
\bibinfo{author}{\bibfnamefont{X.}~\bibnamefont{Luo}}
  (\bibinfo{collaboration}{STAR}), \bibinfo{journal}{PoS}
  \textbf{\bibinfo{volume}{CPOD2014}}, \bibinfo{pages}{019}
  (\bibinfo{year}{2015}), \eprint{1503.02558}.

\bibitem[{\citenamefont{Aamodt et~al.}(2011)}]{Aamodt:2011mr}
\bibinfo{author}{\bibfnamefont{K.}~\bibnamefont{Aamodt}} \bibnamefont{et~al.}
  (\bibinfo{collaboration}{ALICE}), \bibinfo{journal}{Phys. Lett.}
  \textbf{\bibinfo{volume}{B696}}, \bibinfo{pages}{328} (\bibinfo{year}{2011}),
  \eprint{1012.4035}.

\bibitem[{\citenamefont{Adamczyk et~al.}(2015)}]{Adamczyk:2014mxp}
\bibinfo{author}{\bibfnamefont{L.}~\bibnamefont{Adamczyk}} \bibnamefont{et~al.}
  (\bibinfo{collaboration}{STAR}), \bibinfo{journal}{Phys. Rev.}
  \textbf{\bibinfo{volume}{C92}}, \bibinfo{pages}{014904}
  (\bibinfo{year}{2015}), \eprint{1403.4972}.

\bibitem[{\citenamefont{Lacey}(2015)}]{Lacey:2014wqa}
\bibinfo{author}{\bibfnamefont{R.~A.} \bibnamefont{Lacey}},
  \bibinfo{journal}{Phys. Rev. Lett.} \textbf{\bibinfo{volume}{114}},
  \bibinfo{pages}{142301} (\bibinfo{year}{2015}), \eprint{1411.7931}.

\bibitem[{\citenamefont{Zhang}(2019)}]{Zhang:2019wun}
\bibinfo{author}{\bibfnamefont{D.}~\bibnamefont{Zhang}}
  (\bibinfo{collaboration}{STAR}) (\bibinfo{year}{2019}), \eprint{1909.07028}.

\bibitem[{\citenamefont{Karsch and Redlich}(2011)}]{Karsch:2010ck}
\bibinfo{author}{\bibfnamefont{F.}~\bibnamefont{Karsch}} \bibnamefont{and}
  \bibinfo{author}{\bibfnamefont{K.}~\bibnamefont{Redlich}},
  \bibinfo{journal}{Phys. Lett. B} \textbf{\bibinfo{volume}{695}},
  \bibinfo{pages}{136} (\bibinfo{year}{2011}), \eprint{1007.2581}.

\bibitem[{\citenamefont{Borsanyi et~al.}(2013)\citenamefont{Borsanyi, Fodor,
  Katz, Krieg, Ratti, and Szabo}}]{Borsanyi:2013hza}
\bibinfo{author}{\bibfnamefont{S.}~\bibnamefont{Borsanyi}},
  \bibinfo{author}{\bibfnamefont{Z.}~\bibnamefont{Fodor}},
  \bibinfo{author}{\bibfnamefont{S.~D.} \bibnamefont{Katz}},
  \bibinfo{author}{\bibfnamefont{S.}~\bibnamefont{Krieg}},
  \bibinfo{author}{\bibfnamefont{C.}~\bibnamefont{Ratti}}, \bibnamefont{and}
  \bibinfo{author}{\bibfnamefont{K.~K.} \bibnamefont{Szabo}},
  \bibinfo{journal}{Phys. Rev. Lett.} \textbf{\bibinfo{volume}{111}},
  \bibinfo{pages}{062005} (\bibinfo{year}{2013}), \eprint{1305.5161}.

\bibitem[{\citenamefont{Luo et~al.}(2014)\citenamefont{Luo, Mohanty, and
  Xu}}]{Luo:2014tga}
\bibinfo{author}{\bibfnamefont{X.}~\bibnamefont{Luo}},
  \bibinfo{author}{\bibfnamefont{B.}~\bibnamefont{Mohanty}}, \bibnamefont{and}
  \bibinfo{author}{\bibfnamefont{N.}~\bibnamefont{Xu}}, \bibinfo{journal}{Nucl.
  Phys.} \textbf{\bibinfo{volume}{A931}}, \bibinfo{pages}{808}
  (\bibinfo{year}{2014}), \eprint{1408.0495}.

\bibitem[{\citenamefont{Netrakanti et~al.}(2016)\citenamefont{Netrakanti, Luo,
  Mishra, Mohanty, Mohanty, and Xu}}]{Netrakanti:2014mta}
\bibinfo{author}{\bibfnamefont{P.~K.} \bibnamefont{Netrakanti}},
  \bibinfo{author}{\bibfnamefont{X.~F.} \bibnamefont{Luo}},
  \bibinfo{author}{\bibfnamefont{D.~K.} \bibnamefont{Mishra}},
  \bibinfo{author}{\bibfnamefont{B.}~\bibnamefont{Mohanty}},
  \bibinfo{author}{\bibfnamefont{A.}~\bibnamefont{Mohanty}}, \bibnamefont{and}
  \bibinfo{author}{\bibfnamefont{N.}~\bibnamefont{Xu}}, \bibinfo{journal}{Nucl.
  Phys.} \textbf{\bibinfo{volume}{A947}}, \bibinfo{pages}{248}
  (\bibinfo{year}{2016}), \eprint{1405.4617}.

\bibitem[{\citenamefont{Li et~al.}(2018)\citenamefont{Li, Xu, and
  Song}}]{Li:2017via}
\bibinfo{author}{\bibfnamefont{J.}~\bibnamefont{Li}},
  \bibinfo{author}{\bibfnamefont{H.-j.} \bibnamefont{Xu}}, \bibnamefont{and}
  \bibinfo{author}{\bibfnamefont{H.}~\bibnamefont{Song}},
  \bibinfo{journal}{Phys. Rev.} \textbf{\bibinfo{volume}{C97}},
  \bibinfo{pages}{014902} (\bibinfo{year}{2018}), \eprint{1707.09742}.

\bibitem[{\citenamefont{He et~al.}(2016)\citenamefont{He, Luo, Nara, Esumi, and
  Xu}}]{He:2016uei}
\bibinfo{author}{\bibfnamefont{S.}~\bibnamefont{He}},
  \bibinfo{author}{\bibfnamefont{X.}~\bibnamefont{Luo}},
  \bibinfo{author}{\bibfnamefont{Y.}~\bibnamefont{Nara}},
  \bibinfo{author}{\bibfnamefont{S.}~\bibnamefont{Esumi}}, \bibnamefont{and}
  \bibinfo{author}{\bibfnamefont{N.}~\bibnamefont{Xu}}, \bibinfo{journal}{Phys.
  Lett. B} \textbf{\bibinfo{volume}{762}}, \bibinfo{pages}{296}
  (\bibinfo{year}{2016}), \eprint{1607.06376}.

\bibitem[{\citenamefont{Xu et~al.}(2016)\citenamefont{Xu, Yu, Liu, and
  Luo}}]{Xu:2016qjd}
\bibinfo{author}{\bibfnamefont{J.}~\bibnamefont{Xu}},
  \bibinfo{author}{\bibfnamefont{S.}~\bibnamefont{Yu}},
  \bibinfo{author}{\bibfnamefont{F.}~\bibnamefont{Liu}}, \bibnamefont{and}
  \bibinfo{author}{\bibfnamefont{X.}~\bibnamefont{Luo}},
  \bibinfo{journal}{Phys. Rev.} \textbf{\bibinfo{volume}{C94}},
  \bibinfo{pages}{024901} (\bibinfo{year}{2016}), \eprint{1606.03900}.

\bibitem[{\citenamefont{Liu et~al.}(2019)\citenamefont{Liu, Zhang, He, Yu, and
  Luo}}]{Liu:2019nii}
\bibinfo{author}{\bibfnamefont{H.}~\bibnamefont{Liu}},
  \bibinfo{author}{\bibfnamefont{D.}~\bibnamefont{Zhang}},
  \bibinfo{author}{\bibfnamefont{S.}~\bibnamefont{He}},
  \bibinfo{author}{\bibfnamefont{N.}~\bibnamefont{Yu}}, \bibnamefont{and}
  \bibinfo{author}{\bibfnamefont{X.}~\bibnamefont{Luo}} (\bibinfo{year}{2019}),
  \eprint{1909.09304}.

\bibitem[{\citenamefont{Pratt and Plumberg}(2019)}]{Pratt:2018ebf}
\bibinfo{author}{\bibfnamefont{S.}~\bibnamefont{Pratt}} \bibnamefont{and}
  \bibinfo{author}{\bibfnamefont{C.}~\bibnamefont{Plumberg}},
  \bibinfo{journal}{Phys. Rev. C} \textbf{\bibinfo{volume}{99}},
  \bibinfo{pages}{044916} (\bibinfo{year}{2019}), \eprint{1812.05649}.

\bibitem[{\citenamefont{Oliinychenko and Koch}(2019)}]{Oliinychenko:2019zfk}
\bibinfo{author}{\bibfnamefont{D.}~\bibnamefont{Oliinychenko}}
  \bibnamefont{and} \bibinfo{author}{\bibfnamefont{V.}~\bibnamefont{Koch}},
  \bibinfo{journal}{Phys. Rev. Lett.} \textbf{\bibinfo{volume}{123}},
  \bibinfo{pages}{182302} (\bibinfo{year}{2019}), \eprint{1902.09775}.

\bibitem[{\citenamefont{Adamczyk et~al.}(2016)}]{Adamczyk:2016gfs}
\bibinfo{author}{\bibfnamefont{L.}~\bibnamefont{Adamczyk}} \bibnamefont{et~al.}
  (\bibinfo{collaboration}{STAR}), \bibinfo{journal}{Phys. Rev.}
  \textbf{\bibinfo{volume}{C94}}, \bibinfo{pages}{034908}
  (\bibinfo{year}{2016}), \eprint{1601.07052}.

\bibitem[{\citenamefont{Chen et~al.}(2018)\citenamefont{Chen, Keane, Ma, Tang,
  and Xu}}]{Chen:2018tnh}
\bibinfo{author}{\bibfnamefont{J.}~\bibnamefont{Chen}},
  \bibinfo{author}{\bibfnamefont{D.}~\bibnamefont{Keane}},
  \bibinfo{author}{\bibfnamefont{Y.-G.} \bibnamefont{Ma}},
  \bibinfo{author}{\bibfnamefont{A.}~\bibnamefont{Tang}}, \bibnamefont{and}
  \bibinfo{author}{\bibfnamefont{Z.}~\bibnamefont{Xu}}, \bibinfo{journal}{Phys.
  Rept.} \textbf{\bibinfo{volume}{760}}, \bibinfo{pages}{1}
  (\bibinfo{year}{2018}), \eprint{1808.09619}.

\bibitem[{\citenamefont{Adam et~al.}(2019)}]{Adam:2019wnb}
\bibinfo{author}{\bibfnamefont{J.}~\bibnamefont{Adam}} \bibnamefont{et~al.}
  (\bibinfo{collaboration}{STAR}), \bibinfo{journal}{Phys. Rev.}
  \textbf{\bibinfo{volume}{C99}}, \bibinfo{pages}{064905}
  (\bibinfo{year}{2019}), \eprint{1903.11778}.

\bibitem[{\citenamefont{Sun et~al.}(2017)\citenamefont{Sun, Chen, Ko, and
  Xu}}]{Sun:2017xrx}
\bibinfo{author}{\bibfnamefont{K.-J.} \bibnamefont{Sun}},
  \bibinfo{author}{\bibfnamefont{L.-W.} \bibnamefont{Chen}},
  \bibinfo{author}{\bibfnamefont{C.~M.} \bibnamefont{Ko}}, \bibnamefont{and}
  \bibinfo{author}{\bibfnamefont{Z.}~\bibnamefont{Xu}}, \bibinfo{journal}{Phys.
  Lett.} \textbf{\bibinfo{volume}{B774}}, \bibinfo{pages}{103}
  (\bibinfo{year}{2017}), \eprint{1702.07620}.

\bibitem[{\citenamefont{Sun et~al.}(2018)\citenamefont{Sun, Chen, Ko, Pu, and
  Xu}}]{Sun:2018jhg}
\bibinfo{author}{\bibfnamefont{K.-J.} \bibnamefont{Sun}},
  \bibinfo{author}{\bibfnamefont{L.-W.} \bibnamefont{Chen}},
  \bibinfo{author}{\bibfnamefont{C.~M.} \bibnamefont{Ko}},
  \bibinfo{author}{\bibfnamefont{J.}~\bibnamefont{Pu}}, \bibnamefont{and}
  \bibinfo{author}{\bibfnamefont{Z.}~\bibnamefont{Xu}}, \bibinfo{journal}{Phys.
  Lett.} \textbf{\bibinfo{volume}{B781}}, \bibinfo{pages}{499}
  (\bibinfo{year}{2018}), \eprint{1801.09382}.

\bibitem[{\citenamefont{Yu et~al.}(2020)\citenamefont{Yu, Zhang, and
  Luo}}]{Yu:2018kvh}
\bibinfo{author}{\bibfnamefont{N.}~\bibnamefont{Yu}},
  \bibinfo{author}{\bibfnamefont{D.}~\bibnamefont{Zhang}}, \bibnamefont{and}
  \bibinfo{author}{\bibfnamefont{X.}~\bibnamefont{Luo}},
  \bibinfo{journal}{Chin. Phys.} \textbf{\bibinfo{volume}{C44}},
  \bibinfo{pages}{014002} (\bibinfo{year}{2020}), \eprint{1812.04291}.

\bibitem[{\citenamefont{Sun and Ko}(2020)}]{Sun:2020uoj}
\bibinfo{author}{\bibfnamefont{K.-J.} \bibnamefont{Sun}} \bibnamefont{and}
  \bibinfo{author}{\bibfnamefont{C.~M.} \bibnamefont{Ko}}
  (\bibinfo{year}{2020}), \eprint{2005.00182}.

\bibitem[{\citenamefont{Deng and Ma}(2020)}]{Deng:2020zxo}
\bibinfo{author}{\bibfnamefont{X.~G.} \bibnamefont{Deng}} \bibnamefont{and}
  \bibinfo{author}{\bibfnamefont{Y.~G.} \bibnamefont{Ma}}
  (\bibinfo{year}{2020}), \eprint{2006.12337}.

\bibitem[{\citenamefont{Shen and Schenke}(2018{\natexlab{a}})}]{Shen:2017bsr}
\bibinfo{author}{\bibfnamefont{C.}~\bibnamefont{Shen}} \bibnamefont{and}
  \bibinfo{author}{\bibfnamefont{B.}~\bibnamefont{Schenke}},
  \bibinfo{journal}{Phys. Rev.} \textbf{\bibinfo{volume}{C97}},
  \bibinfo{pages}{024907} (\bibinfo{year}{2018}{\natexlab{a}}),
  \eprint{1710.00881}.

\bibitem[{\citenamefont{Shen and Schenke}(2018{\natexlab{b}})}]{Shen:2017fnn}
\bibinfo{author}{\bibfnamefont{C.}~\bibnamefont{Shen}} \bibnamefont{and}
  \bibinfo{author}{\bibfnamefont{B.}~\bibnamefont{Schenke}},
  \bibinfo{journal}{PoS} \textbf{\bibinfo{volume}{CPOD2017}},
  \bibinfo{pages}{006} (\bibinfo{year}{2018}{\natexlab{b}}),
  \eprint{1711.10544}.

\bibitem[{3DG()}]{3DGlauber}
\bibinfo{note}{The 3D Monte-Carlo Glauber Model is a code package to simulate
  high-energy nucleus-nucleus collisions in 3D. The longitudinal
  energy-momentum distribution is based on the classical string deceleration
  model. This work uses v0.5 of this framework, which can be downloaded from
  \url{https://github.com/chunshen1987/3dMCGlauber}.}

\bibitem[{\citenamefont{Denicol et~al.}(2018)\citenamefont{Denicol, Gale, Jeon,
  Monnai, Schenke, and Shen}}]{Denicol:2018wdp}
\bibinfo{author}{\bibfnamefont{G.~S.} \bibnamefont{Denicol}},
  \bibinfo{author}{\bibfnamefont{C.}~\bibnamefont{Gale}},
  \bibinfo{author}{\bibfnamefont{S.}~\bibnamefont{Jeon}},
  \bibinfo{author}{\bibfnamefont{A.}~\bibnamefont{Monnai}},
  \bibinfo{author}{\bibfnamefont{B.}~\bibnamefont{Schenke}}, \bibnamefont{and}
  \bibinfo{author}{\bibfnamefont{C.}~\bibnamefont{Shen}},
  \bibinfo{journal}{Phys. Rev.} \textbf{\bibinfo{volume}{C98}},
  \bibinfo{pages}{034916} (\bibinfo{year}{2018}), \eprint{1804.10557}.

\bibitem[{\citenamefont{Shen and Schenke}(2019)}]{Shen:2018pty}
\bibinfo{author}{\bibfnamefont{C.}~\bibnamefont{Shen}} \bibnamefont{and}
  \bibinfo{author}{\bibfnamefont{B.}~\bibnamefont{Schenke}},
  \bibinfo{journal}{Nucl. Phys.} \textbf{\bibinfo{volume}{A982}},
  \bibinfo{pages}{411} (\bibinfo{year}{2019}), \eprint{1807.05141}.

\bibitem[{\citenamefont{Mattiello et~al.}(1995)\citenamefont{Mattiello, Jahns,
  Sorge, Stoecker, and Greiner}}]{Mattiello:1995xg}
\bibinfo{author}{\bibfnamefont{R.}~\bibnamefont{Mattiello}},
  \bibinfo{author}{\bibfnamefont{A.}~\bibnamefont{Jahns}},
  \bibinfo{author}{\bibfnamefont{H.}~\bibnamefont{Sorge}},
  \bibinfo{author}{\bibfnamefont{H.}~\bibnamefont{Stoecker}}, \bibnamefont{and}
  \bibinfo{author}{\bibfnamefont{W.}~\bibnamefont{Greiner}},
  \bibinfo{journal}{Phys. Rev. Lett.} \textbf{\bibinfo{volume}{74}},
  \bibinfo{pages}{2180} (\bibinfo{year}{1995}).

\bibitem[{\citenamefont{Mattiello et~al.}(1997)\citenamefont{Mattiello, Sorge,
  Stoecker, and Greiner}}]{Mattiello:1996gq}
\bibinfo{author}{\bibfnamefont{R.}~\bibnamefont{Mattiello}},
  \bibinfo{author}{\bibfnamefont{H.}~\bibnamefont{Sorge}},
  \bibinfo{author}{\bibfnamefont{H.}~\bibnamefont{Stoecker}}, \bibnamefont{and}
  \bibinfo{author}{\bibfnamefont{W.}~\bibnamefont{Greiner}},
  \bibinfo{journal}{Phys. Rev.} \textbf{\bibinfo{volume}{C55}},
  \bibinfo{pages}{1443} (\bibinfo{year}{1997}), \eprint{nucl-th/9607003}.

\bibitem[{\citenamefont{Chen et~al.}(2003{\natexlab{a}})\citenamefont{Chen, Ko,
  and Li}}]{Chen:2003qj}
\bibinfo{author}{\bibfnamefont{L.-W.} \bibnamefont{Chen}},
  \bibinfo{author}{\bibfnamefont{C.~M.} \bibnamefont{Ko}}, \bibnamefont{and}
  \bibinfo{author}{\bibfnamefont{B.-A.} \bibnamefont{Li}},
  \bibinfo{journal}{Phys. Rev.} \textbf{\bibinfo{volume}{C68}},
  \bibinfo{pages}{017601} (\bibinfo{year}{2003}{\natexlab{a}}),
  \eprint{nucl-th/0302068}.

\bibitem[{\citenamefont{Chen et~al.}(2003{\natexlab{b}})\citenamefont{Chen, Ko,
  and Li}}]{Chen:2003ava}
\bibinfo{author}{\bibfnamefont{L.-W.} \bibnamefont{Chen}},
  \bibinfo{author}{\bibfnamefont{C.~M.} \bibnamefont{Ko}}, \bibnamefont{and}
  \bibinfo{author}{\bibfnamefont{B.-A.} \bibnamefont{Li}},
  \bibinfo{journal}{Nucl. Phys.} \textbf{\bibinfo{volume}{A729}},
  \bibinfo{pages}{809} (\bibinfo{year}{2003}{\natexlab{b}}),
  \eprint{nucl-th/0306032}.

\bibitem[{\citenamefont{Ko et~al.}(2014)\citenamefont{Ko, Song, Li, Greco, and
  Plumari}}]{Song:2012cd}
\bibinfo{author}{\bibfnamefont{C.~M.} \bibnamefont{Ko}},
  \bibinfo{author}{\bibfnamefont{T.}~\bibnamefont{Song}},
  \bibinfo{author}{\bibfnamefont{F.}~\bibnamefont{Li}},
  \bibinfo{author}{\bibfnamefont{V.}~\bibnamefont{Greco}}, \bibnamefont{and}
  \bibinfo{author}{\bibfnamefont{S.}~\bibnamefont{Plumari}},
  \bibinfo{journal}{Nucl. Phys.} \textbf{\bibinfo{volume}{A928}},
  \bibinfo{pages}{234} (\bibinfo{year}{2014}), \eprint{1211.5511}.

\bibitem[{\citenamefont{Angeli and Marinova}(2013)}]{Angeli:2013epw}
\bibinfo{author}{\bibfnamefont{I.}~\bibnamefont{Angeli}} \bibnamefont{and}
  \bibinfo{author}{\bibfnamefont{K.~P.} \bibnamefont{Marinova}},
  \bibinfo{journal}{Atom. Data Nucl. Data Tabl.} \textbf{\bibinfo{volume}{99}},
  \bibinfo{pages}{69} (\bibinfo{year}{2013}).

\bibitem[{iEB()}]{iEBEMUSIC}
\bibinfo{note}{The iEBE-MUSIC is a general-purpose numerical framework to
  simulate dynamical evolution of relativistic heavy-ion collisions
  event-by-event. This work uses v0.5 of this framework, which can be
  downloaded from \url{https://github.com/chunshen1987/iEBE-MUSIC}.}

\bibitem[{\citenamefont{Schenke et~al.}(2011)\citenamefont{Schenke, Jeon, and
  Gale}}]{Schenke:2010rr}
\bibinfo{author}{\bibfnamefont{B.}~\bibnamefont{Schenke}},
  \bibinfo{author}{\bibfnamefont{S.}~\bibnamefont{Jeon}}, \bibnamefont{and}
  \bibinfo{author}{\bibfnamefont{C.}~\bibnamefont{Gale}},
  \bibinfo{journal}{Phys. Rev. Lett.} \textbf{\bibinfo{volume}{106}},
  \bibinfo{pages}{042301} (\bibinfo{year}{2011}), \eprint{1009.3244}.

\bibitem[{\citenamefont{Schenke et~al.}(2010)\citenamefont{Schenke, Jeon, and
  Gale}}]{Schenke:2010nt}
\bibinfo{author}{\bibfnamefont{B.}~\bibnamefont{Schenke}},
  \bibinfo{author}{\bibfnamefont{S.}~\bibnamefont{Jeon}}, \bibnamefont{and}
  \bibinfo{author}{\bibfnamefont{C.}~\bibnamefont{Gale}},
  \bibinfo{journal}{Phys. Rev.} \textbf{\bibinfo{volume}{C82}},
  \bibinfo{pages}{014903} (\bibinfo{year}{2010}), \eprint{1004.1408}.

\bibitem[{\citenamefont{Paquet et~al.}(2016)\citenamefont{Paquet, Shen,
  Denicol, Luzum, Schenke, Jeon, and Gale}}]{Paquet:2015lta}
\bibinfo{author}{\bibfnamefont{J.-F.} \bibnamefont{Paquet}},
  \bibinfo{author}{\bibfnamefont{C.}~\bibnamefont{Shen}},
  \bibinfo{author}{\bibfnamefont{G.~S.} \bibnamefont{Denicol}},
  \bibinfo{author}{\bibfnamefont{M.}~\bibnamefont{Luzum}},
  \bibinfo{author}{\bibfnamefont{B.}~\bibnamefont{Schenke}},
  \bibinfo{author}{\bibfnamefont{S.}~\bibnamefont{Jeon}}, \bibnamefont{and}
  \bibinfo{author}{\bibfnamefont{C.}~\bibnamefont{Gale}},
  \bibinfo{journal}{Phys. Rev. C} \textbf{\bibinfo{volume}{93}},
  \bibinfo{pages}{044906} (\bibinfo{year}{2016}), \eprint{1509.06738}.

\bibitem[{MUS()}]{MUSIC}
\bibinfo{note}{MUSIC is the numerical implementation of (3+1)D relativistic
  viscous hydrodynamic simulations for high energy heavy-ion collisions. Its
  official website is \url{http: //www.physics.mcgill.ca/music}. This work uses
  v2.5 of this framework, which can be downloaded from
  \url{https://github.com/MUSIC-fluid/MUSIC}.}

\bibitem[{\citenamefont{Bass et~al.}(1998)}]{Bass:1998ca}
\bibinfo{author}{\bibfnamefont{S.~A.} \bibnamefont{Bass}} \bibnamefont{et~al.},
  \bibinfo{journal}{Prog. Part. Nucl. Phys.} \textbf{\bibinfo{volume}{41}},
  \bibinfo{pages}{255} (\bibinfo{year}{1998}), \bibinfo{note}{[Prog. Part.
  Nucl. Phys.41,225(1998)]}, \eprint{nucl-th/9803035}.

\bibitem[{\citenamefont{Bleicher et~al.}(1999)}]{Bleicher:1999xi}
\bibinfo{author}{\bibfnamefont{M.}~\bibnamefont{Bleicher}}
  \bibnamefont{et~al.}, \bibinfo{journal}{J. Phys.}
  \textbf{\bibinfo{volume}{G25}}, \bibinfo{pages}{1859} (\bibinfo{year}{1999}),
  \eprint{hep-ph/9909407}.

\bibitem[{UrQ()}]{UrQMD}
\bibinfo{note}{We use the official UrQMD v3.4 and set it up to run as the
  afterburner mode, \url{https://bitbucket.org/
  Chunshen1987/urqmd_afterburner/src/master/}.}

\bibitem[{\citenamefont{Bialas et~al.}(2018)\citenamefont{Bialas, Bzdak, and
  Koch}}]{Bialas:2016epd}
\bibinfo{author}{\bibfnamefont{A.}~\bibnamefont{Bialas}},
  \bibinfo{author}{\bibfnamefont{A.}~\bibnamefont{Bzdak}}, \bibnamefont{and}
  \bibinfo{author}{\bibfnamefont{V.}~\bibnamefont{Koch}},
  \bibinfo{journal}{Acta Phys. Polon.} \textbf{\bibinfo{volume}{B49}},
  \bibinfo{pages}{103} (\bibinfo{year}{2018}), \eprint{1608.07041}.

\bibitem[{\citenamefont{Kharzeev}(1996)}]{Kharzeev:1996sq}
\bibinfo{author}{\bibfnamefont{D.}~\bibnamefont{Kharzeev}},
  \bibinfo{journal}{Phys. Lett.} \textbf{\bibinfo{volume}{B378}},
  \bibinfo{pages}{238} (\bibinfo{year}{1996}), \eprint{nucl-th/9602027}.

\bibitem[{\citenamefont{Shen and Schenke}()}]{Shenfuture1}
\bibinfo{author}{\bibfnamefont{C.}~\bibnamefont{Shen}} \bibnamefont{and}
  \bibinfo{author}{\bibfnamefont{B.}~\bibnamefont{Schenke}}, \bibinfo{note}{in
  preparation.}

\bibitem[{\citenamefont{Borsanyi et~al.}(2012)\citenamefont{Borsanyi, Fodor,
  Katz, Krieg, Ratti, and Szabo}}]{Borsanyi:2011sw}
\bibinfo{author}{\bibfnamefont{S.}~\bibnamefont{Borsanyi}},
  \bibinfo{author}{\bibfnamefont{Z.}~\bibnamefont{Fodor}},
  \bibinfo{author}{\bibfnamefont{S.~D.} \bibnamefont{Katz}},
  \bibinfo{author}{\bibfnamefont{S.}~\bibnamefont{Krieg}},
  \bibinfo{author}{\bibfnamefont{C.}~\bibnamefont{Ratti}}, \bibnamefont{and}
  \bibinfo{author}{\bibfnamefont{K.}~\bibnamefont{Szabo}},
  \bibinfo{journal}{JHEP} \textbf{\bibinfo{volume}{01}}, \bibinfo{pages}{138}
  (\bibinfo{year}{2012}), \eprint{1112.4416}.

\bibitem[{\citenamefont{Borsanyi et~al.}(2014)\citenamefont{Borsanyi, Fodor,
  Hoelbling, Katz, Krieg, and Szabo}}]{Borsanyi:2013bia}
\bibinfo{author}{\bibfnamefont{S.}~\bibnamefont{Borsanyi}},
  \bibinfo{author}{\bibfnamefont{Z.}~\bibnamefont{Fodor}},
  \bibinfo{author}{\bibfnamefont{C.}~\bibnamefont{Hoelbling}},
  \bibinfo{author}{\bibfnamefont{S.~D.} \bibnamefont{Katz}},
  \bibinfo{author}{\bibfnamefont{S.}~\bibnamefont{Krieg}}, \bibnamefont{and}
  \bibinfo{author}{\bibfnamefont{K.~K.} \bibnamefont{Szabo}},
  \bibinfo{journal}{Phys. Lett.} \textbf{\bibinfo{volume}{B730}},
  \bibinfo{pages}{99} (\bibinfo{year}{2014}), \eprint{1309.5258}.

\bibitem[{\citenamefont{Ding et~al.}(2015)\citenamefont{Ding, Mukherjee, Ohno,
  Petreczky, and Schadler}}]{Ding:2015fca}
\bibinfo{author}{\bibfnamefont{H.~T.} \bibnamefont{Ding}},
  \bibinfo{author}{\bibfnamefont{S.}~\bibnamefont{Mukherjee}},
  \bibinfo{author}{\bibfnamefont{H.}~\bibnamefont{Ohno}},
  \bibinfo{author}{\bibfnamefont{P.}~\bibnamefont{Petreczky}},
  \bibnamefont{and} \bibinfo{author}{\bibfnamefont{H.~P.}
  \bibnamefont{Schadler}}, \bibinfo{journal}{Phys. Rev. D}
  \textbf{\bibinfo{volume}{92}}, \bibinfo{pages}{074043}
  (\bibinfo{year}{2015}), \eprint{1507.06637}.

\bibitem[{\citenamefont{Bazavov et~al.}(2017)}]{Bazavov:2017dus}
\bibinfo{author}{\bibfnamefont{A.}~\bibnamefont{Bazavov}} \bibnamefont{et~al.},
  \bibinfo{journal}{Phys. Rev. D} \textbf{\bibinfo{volume}{95}},
  \bibinfo{pages}{054504} (\bibinfo{year}{2017}), \eprint{1701.04325}.

\bibitem[{\citenamefont{Monnai et~al.}(2019)\citenamefont{Monnai, Schenke, and
  Shen}}]{Monnai:2019hkn}
\bibinfo{author}{\bibfnamefont{A.}~\bibnamefont{Monnai}},
  \bibinfo{author}{\bibfnamefont{B.}~\bibnamefont{Schenke}}, \bibnamefont{and}
  \bibinfo{author}{\bibfnamefont{C.}~\bibnamefont{Shen}},
  \bibinfo{journal}{Phys. Rev.} \textbf{\bibinfo{volume}{C100}},
  \bibinfo{pages}{024907} (\bibinfo{year}{2019}), \eprint{1902.05095}.

\bibitem[{\citenamefont{Denicol et~al.}(2012)\citenamefont{Denicol, Niemi,
  Molnar, and Rischke}}]{Denicol:2012cn}
\bibinfo{author}{\bibfnamefont{G.}~\bibnamefont{Denicol}},
  \bibinfo{author}{\bibfnamefont{H.}~\bibnamefont{Niemi}},
  \bibinfo{author}{\bibfnamefont{E.}~\bibnamefont{Molnar}}, \bibnamefont{and}
  \bibinfo{author}{\bibfnamefont{D.}~\bibnamefont{Rischke}},
  \bibinfo{journal}{Phys. Rev. D} \textbf{\bibinfo{volume}{85}},
  \bibinfo{pages}{114047} (\bibinfo{year}{2012}), \bibinfo{note}{[Erratum:
  Phys.Rev.D 91, 039902 (2015)]}, \eprint{1202.4551}.

\bibitem[{\citenamefont{Shen et~al.}(2016)\citenamefont{Shen, Qiu, Song,
  Bernhard, Bass, and Heinz}}]{Shen:2014vra}
\bibinfo{author}{\bibfnamefont{C.}~\bibnamefont{Shen}},
  \bibinfo{author}{\bibfnamefont{Z.}~\bibnamefont{Qiu}},
  \bibinfo{author}{\bibfnamefont{H.}~\bibnamefont{Song}},
  \bibinfo{author}{\bibfnamefont{J.}~\bibnamefont{Bernhard}},
  \bibinfo{author}{\bibfnamefont{S.}~\bibnamefont{Bass}}, \bibnamefont{and}
  \bibinfo{author}{\bibfnamefont{U.}~\bibnamefont{Heinz}},
  \bibinfo{journal}{Comput. Phys. Commun.} \textbf{\bibinfo{volume}{199}},
  \bibinfo{pages}{61} (\bibinfo{year}{2016}), \eprint{1409.8164}.

\bibitem[{iSS()}]{iSS_code}
\bibinfo{note}{The iSS code package is an open-source particle sampler based on
  the Cooper-Frye freeze-out prescription. It converts fluid cells to particle
  samples. This work uses v1.0 of the iSS, which can be downloaded from
  \url{https://github.com/chunshen1987/iSS/releases}.}

\bibitem[{\citenamefont{Adamczyk et~al.}(2018)}]{Adamczyk:2017nof}
\bibinfo{author}{\bibfnamefont{L.}~\bibnamefont{Adamczyk}} \bibnamefont{et~al.}
  (\bibinfo{collaboration}{STAR}), \bibinfo{journal}{Phys. Rev. Lett.}
  \textbf{\bibinfo{volume}{121}}, \bibinfo{pages}{032301}
  (\bibinfo{year}{2018}), \eprint{1707.01988}.

\bibitem[{\citenamefont{Adler et~al.}(2004)}]{Adler:2003cb}
\bibinfo{author}{\bibfnamefont{S.~S.} \bibnamefont{Adler}} \bibnamefont{et~al.}
  (\bibinfo{collaboration}{PHENIX}), \bibinfo{journal}{Phys. Rev.}
  \textbf{\bibinfo{volume}{C69}}, \bibinfo{pages}{034909}
  (\bibinfo{year}{2004}), \eprint{nucl-ex/0307022}.

\bibitem[{\citenamefont{Zhao et~al.}(2017)\citenamefont{Zhao, Xu, and
  Song}}]{Zhao:2017yhj}
\bibinfo{author}{\bibfnamefont{W.}~\bibnamefont{Zhao}},
  \bibinfo{author}{\bibfnamefont{H.-j.} \bibnamefont{Xu}}, \bibnamefont{and}
  \bibinfo{author}{\bibfnamefont{H.}~\bibnamefont{Song}},
  \bibinfo{journal}{Eur. Phys. J.} \textbf{\bibinfo{volume}{C77}},
  \bibinfo{pages}{645} (\bibinfo{year}{2017}), \eprint{1703.10792}.

\bibitem[{\citenamefont{Zhao et~al.}(2018)\citenamefont{Zhao, Zhu, Zheng, Ko,
  and Song}}]{Zhao:2018lyf}
\bibinfo{author}{\bibfnamefont{W.}~\bibnamefont{Zhao}},
  \bibinfo{author}{\bibfnamefont{L.}~\bibnamefont{Zhu}},
  \bibinfo{author}{\bibfnamefont{H.}~\bibnamefont{Zheng}},
  \bibinfo{author}{\bibfnamefont{C.~M.} \bibnamefont{Ko}}, \bibnamefont{and}
  \bibinfo{author}{\bibfnamefont{H.}~\bibnamefont{Song}},
  \bibinfo{journal}{Phys. Rev.} \textbf{\bibinfo{volume}{C98}},
  \bibinfo{pages}{054905} (\bibinfo{year}{2018}), \eprint{1807.02813}.

\bibitem[{\citenamefont{Shen}(2020)}]{Shen:2020gef}
\bibinfo{author}{\bibfnamefont{C.}~\bibnamefont{Shen}} (\bibinfo{year}{2020}),
  \eprint{2001.11858}.

\bibitem[{\citenamefont{Greco et~al.}(2003)\citenamefont{Greco, Ko, and
  Levai}}]{Greco:2003mm}
\bibinfo{author}{\bibfnamefont{V.}~\bibnamefont{Greco}},
  \bibinfo{author}{\bibfnamefont{C.~M.} \bibnamefont{Ko}}, \bibnamefont{and}
  \bibinfo{author}{\bibfnamefont{P.}~\bibnamefont{Levai}},
  \bibinfo{journal}{Phys. Rev.} \textbf{\bibinfo{volume}{C68}},
  \bibinfo{pages}{034904} (\bibinfo{year}{2003}), \eprint{nucl-th/0305024}.

\bibitem[{\citenamefont{Greco et~al.}(2004)\citenamefont{Greco, Ko, and
  Rapp}}]{Greco:2003vf}
\bibinfo{author}{\bibfnamefont{V.}~\bibnamefont{Greco}},
  \bibinfo{author}{\bibfnamefont{C.~M.} \bibnamefont{Ko}}, \bibnamefont{and}
  \bibinfo{author}{\bibfnamefont{R.}~\bibnamefont{Rapp}},
  \bibinfo{journal}{Phys. Lett.} \textbf{\bibinfo{volume}{B595}},
  \bibinfo{pages}{202} (\bibinfo{year}{2004}), \eprint{nucl-th/0312100}.

\bibitem[{\citenamefont{Fries et~al.}(2003)\citenamefont{Fries, Muller, Nonaka,
  and Bass}}]{Fries:2003kq}
\bibinfo{author}{\bibfnamefont{R.~J.} \bibnamefont{Fries}},
  \bibinfo{author}{\bibfnamefont{B.}~\bibnamefont{Muller}},
  \bibinfo{author}{\bibfnamefont{C.}~\bibnamefont{Nonaka}}, \bibnamefont{and}
  \bibinfo{author}{\bibfnamefont{S.~A.} \bibnamefont{Bass}},
  \bibinfo{journal}{Phys. Rev.} \textbf{\bibinfo{volume}{C68}},
  \bibinfo{pages}{044902} (\bibinfo{year}{2003}), \eprint{nucl-th/0306027}.

\bibitem[{\citenamefont{Hwa and Yang}(2004)}]{Hwa:2004ng}
\bibinfo{author}{\bibfnamefont{R.~C.} \bibnamefont{Hwa}} \bibnamefont{and}
  \bibinfo{author}{\bibfnamefont{C.~B.} \bibnamefont{Yang}},
  \bibinfo{journal}{Phys. Rev.} \textbf{\bibinfo{volume}{C70}},
  \bibinfo{pages}{024905} (\bibinfo{year}{2004}), \eprint{nucl-th/0401001}.

\bibitem[{\citenamefont{Fries et~al.}(2008)\citenamefont{Fries, Greco, and
  Sorensen}}]{Fries:2008hs}
\bibinfo{author}{\bibfnamefont{R.~J.} \bibnamefont{Fries}},
  \bibinfo{author}{\bibfnamefont{V.}~\bibnamefont{Greco}}, \bibnamefont{and}
  \bibinfo{author}{\bibfnamefont{P.}~\bibnamefont{Sorensen}},
  \bibinfo{journal}{Ann. Rev. Nucl. Part. Sci.} \textbf{\bibinfo{volume}{58}},
  \bibinfo{pages}{177} (\bibinfo{year}{2008}), \eprint{0807.4939}.

\bibitem[{\citenamefont{Zhao et~al.}(2019)\citenamefont{Zhao, Ko, Liu, Qin, and
  Song}}]{Zhao:2019ehg}
\bibinfo{author}{\bibfnamefont{W.}~\bibnamefont{Zhao}},
  \bibinfo{author}{\bibfnamefont{C.~M.} \bibnamefont{Ko}},
  \bibinfo{author}{\bibfnamefont{Y.-X.} \bibnamefont{Liu}},
  \bibinfo{author}{\bibfnamefont{G.-Y.} \bibnamefont{Qin}}, \bibnamefont{and}
  \bibinfo{author}{\bibfnamefont{H.}~\bibnamefont{Song}}
  (\bibinfo{year}{2019}), \eprint{1911.00826}.

\bibitem[{\citenamefont{Andronic et~al.}(2018)\citenamefont{Andronic,
  Braun-Munzinger, Redlich, and Stachel}}]{Andronic:2017pug}
\bibinfo{author}{\bibfnamefont{A.}~\bibnamefont{Andronic}},
  \bibinfo{author}{\bibfnamefont{P.}~\bibnamefont{Braun-Munzinger}},
  \bibinfo{author}{\bibfnamefont{K.}~\bibnamefont{Redlich}}, \bibnamefont{and}
  \bibinfo{author}{\bibfnamefont{J.}~\bibnamefont{Stachel}},
  \bibinfo{journal}{Nature} \textbf{\bibinfo{volume}{561}},
  \bibinfo{pages}{321} (\bibinfo{year}{2018}), \eprint{1710.09425}.

\bibitem[{\citenamefont{Sun et~al.}(2020)\citenamefont{Sun, Ko, Li, Xu, and
  Chen}}]{Sun:2020pjz}
\bibinfo{author}{\bibfnamefont{K.-J.} \bibnamefont{Sun}},
  \bibinfo{author}{\bibfnamefont{C.~M.} \bibnamefont{Ko}},
  \bibinfo{author}{\bibfnamefont{F.}~\bibnamefont{Li}},
  \bibinfo{author}{\bibfnamefont{J.}~\bibnamefont{Xu}}, \bibnamefont{and}
  \bibinfo{author}{\bibfnamefont{L.-W.} \bibnamefont{Chen}}
  (\bibinfo{year}{2020}), \eprint{2006.08929}.

\bibitem[{\citenamefont{Csernai and Kapusta}(1986)}]{Csernai:1986qf}
\bibinfo{author}{\bibfnamefont{L.~P.} \bibnamefont{Csernai}} \bibnamefont{and}
  \bibinfo{author}{\bibfnamefont{J.~I.} \bibnamefont{Kapusta}},
  \bibinfo{journal}{Phys. Rept.} \textbf{\bibinfo{volume}{131}},
  \bibinfo{pages}{223} (\bibinfo{year}{1986}).

\bibitem[{\citenamefont{Scheibl and Heinz}(1999)}]{Scheibl:1998tk}
\bibinfo{author}{\bibfnamefont{R.}~\bibnamefont{Scheibl}} \bibnamefont{and}
  \bibinfo{author}{\bibfnamefont{U.~W.} \bibnamefont{Heinz}},
  \bibinfo{journal}{Phys. Rev.} \textbf{\bibinfo{volume}{C59}},
  \bibinfo{pages}{1585} (\bibinfo{year}{1999}), \eprint{nucl-th/9809092}.

\bibitem[{\citenamefont{Bellini and Kalweit}(2019)}]{Bellini:2018epz}
\bibinfo{author}{\bibfnamefont{F.}~\bibnamefont{Bellini}} \bibnamefont{and}
  \bibinfo{author}{\bibfnamefont{A.~P.} \bibnamefont{Kalweit}},
  \bibinfo{journal}{Phys. Rev.} \textbf{\bibinfo{volume}{C99}},
  \bibinfo{pages}{054905} (\bibinfo{year}{2019}), \eprint{1807.05894}.

\bibitem[{\citenamefont{Oliinychenko et~al.}(2020)\citenamefont{Oliinychenko,
  Shen, and Koch}}]{Oliinychenko:2020znl}
\bibinfo{author}{\bibfnamefont{D.}~\bibnamefont{Oliinychenko}},
  \bibinfo{author}{\bibfnamefont{C.}~\bibnamefont{Shen}}, \bibnamefont{and}
  \bibinfo{author}{\bibfnamefont{V.}~\bibnamefont{Koch}}
  (\bibinfo{year}{2020}), \eprint{2009.01915}.

\bibitem[{\citenamefont{Shen and Alzhrani}(2020)}]{Shen:2020jwv}
\bibinfo{author}{\bibfnamefont{C.}~\bibnamefont{Shen}} \bibnamefont{and}
  \bibinfo{author}{\bibfnamefont{S.}~\bibnamefont{Alzhrani}},
  \bibinfo{journal}{Phys. Rev. C} \textbf{\bibinfo{volume}{102}},
  \bibinfo{pages}{014909} (\bibinfo{year}{2020}), \eprint{2003.05852}.

\bibitem[{\citenamefont{Shuryak and
  Torres-Rincon}(2019{\natexlab{a}})}]{Shuryak:2018lgd}
\bibinfo{author}{\bibfnamefont{E.}~\bibnamefont{Shuryak}} \bibnamefont{and}
  \bibinfo{author}{\bibfnamefont{J.~M.} \bibnamefont{Torres-Rincon}},
  \bibinfo{journal}{Phys. Rev.} \textbf{\bibinfo{volume}{C100}},
  \bibinfo{pages}{024903} (\bibinfo{year}{2019}{\natexlab{a}}),
  \eprint{1805.04444}.

\bibitem[{\citenamefont{Shuryak and
  Torres-Rincon}(2019{\natexlab{b}})}]{Shuryak:2019ikv}
\bibinfo{author}{\bibfnamefont{E.}~\bibnamefont{Shuryak}} \bibnamefont{and}
  \bibinfo{author}{\bibfnamefont{J.~M.} \bibnamefont{Torres-Rincon}}
  (\bibinfo{year}{2019}{\natexlab{b}}), \eprint{1910.08119}.

\bibitem[{\citenamefont{Sun et~al.}(2019)\citenamefont{Sun, Ko, and
  Donigus}}]{Sun:2018mqq}
\bibinfo{author}{\bibfnamefont{K.-J.} \bibnamefont{Sun}},
  \bibinfo{author}{\bibfnamefont{C.~M.} \bibnamefont{Ko}}, \bibnamefont{and}
  \bibinfo{author}{\bibfnamefont{B.}~\bibnamefont{Donigus}},
  \bibinfo{journal}{Phys. Lett.} \textbf{\bibinfo{volume}{B792}},
  \bibinfo{pages}{132} (\bibinfo{year}{2019}), \eprint{1812.05175}.

\bibitem[{\citenamefont{Nara et~al.}(2000)\citenamefont{Nara, Otuka, Ohnishi,
  Niita, and Chiba}}]{Nara:1999dz}
\bibinfo{author}{\bibfnamefont{Y.}~\bibnamefont{Nara}},
  \bibinfo{author}{\bibfnamefont{N.}~\bibnamefont{Otuka}},
  \bibinfo{author}{\bibfnamefont{A.}~\bibnamefont{Ohnishi}},
  \bibinfo{author}{\bibfnamefont{K.}~\bibnamefont{Niita}}, \bibnamefont{and}
  \bibinfo{author}{\bibfnamefont{S.}~\bibnamefont{Chiba}},
  \bibinfo{journal}{Phys. Rev. C} \textbf{\bibinfo{volume}{61}},
  \bibinfo{pages}{024901} (\bibinfo{year}{2000}), \eprint{nucl-th/9904059}.

\bibitem[{\citenamefont{Lin et~al.}(2005)\citenamefont{Lin, Ko, Li, Zhang, and
  Pal}}]{Lin:2004en}
\bibinfo{author}{\bibfnamefont{Z.-W.} \bibnamefont{Lin}},
  \bibinfo{author}{\bibfnamefont{C.~M.} \bibnamefont{Ko}},
  \bibinfo{author}{\bibfnamefont{B.-A.} \bibnamefont{Li}},
  \bibinfo{author}{\bibfnamefont{B.}~\bibnamefont{Zhang}}, \bibnamefont{and}
  \bibinfo{author}{\bibfnamefont{S.}~\bibnamefont{Pal}},
  \bibinfo{journal}{Phys. Rev. C} \textbf{\bibinfo{volume}{72}},
  \bibinfo{pages}{064901} (\bibinfo{year}{2005}), \eprint{nucl-th/0411110}.

\end{thebibliography}

\end{document}